\documentclass[final,authoryear,1p,times]{elsarticle}

\usepackage{graphicx}

\usepackage{amssymb}

\journal{Theoretical Population Biology}

\begin{document}

\begin{frontmatter}



\title{Growth, competition and cooperation in spatial population genetics}


\author[UPC]{S. Pigolotti \corref{cor1}}
\ead{simone.pigolotti@upc.edu}
\cortext[cor1]{Phone: +34 93 739 8573
Fax: +34 93 739 8000}
\author[TorVergata]{R. Benzi}
\author[Eind]{P. Perlekar}
\author[NBI]{M. H. Jensen }
\author[Eind]{F. Toschi}
\author[Harvard]{D. R. Nelson}

\address[UPC]{Dept. de Fisica i Eng. Nuclear, Universitat Politecnica de Catalunya
  Edif. GAIA, Rambla Sant Nebridi s/n, 08222 Terrassa, Barcelona,
  Spain.}
\address[TorVergata]{Dipartimento di Fisica, Universita' di Roma ``Tor
  Vergata'' and INFN, via della Ricerca Scientifica 1, 00133 Roma,
  Italy.}
\address[Eind]{Department of Physics, Department of Mathematics and
  Computer Science, and J.M. Burgerscentrum, Eindhoven University of
  Technology, 5600 MB Eindhoven, The Netherlands}
\address[NBI]{The Niels Bohr Institut, Blegdamsvej 17, DD-2100 Copenhagen, Denmark.}
\address[Harvard]{Lyman Laboratory of Physics, Harvard University,
  Cambridge, MA 02138, USA}

\begin{abstract}
  We study an individual based model describing competition in space
  between two different alleles. Although the model is similar in
  spirit to classic models of spatial population genetics such as the
  stepping stone model, here however space is continuous and the total
  density of competing individuals fluctuates due to demographic
  stochasticity. By means of analytics and numerical simulations, we
  study the behavior of fixation probabilities, fixation times, and
  heterozygosity, in a neutral setting and in cases where the two
  species can compete or cooperate. By concluding with examples in
  which individuals are transported by fluid flows, we argue that this
  model is a natural choice to describe competition in marine
  environments.
\end{abstract}


\end{frontmatter}

\newpage


\section{Introduction}\label{introduction}

A mathematical analysis of the fate of mutations in spatially extended
populations has been a classic topic of research in population
genetics for at least seventy years
\citep{fisher,kolmogorov,wright_distance,kimuraSSM,kimuraweiss}. 
This interest has
nevertheless increased recently, as improved sequencing technology
allows direct observations of structured genetic diversity in space for
many different species.

On the theoretical side, a landmark in this research has been the
stepping stone model (SSM) proposed by Kimura
\citep{kimuraSSM,kimuraweiss}.  This model considers $m$ islands (or
``demes''), each having a fixed local population size $N_l$ and
arranged along a line or in a regular lattice in more than one spatial
dimension. The population on each island is made up of several species
(or alleles) described by, e.g., a Wright-Fisher or Moran process. Spatial
migration is modeled by assuming that neighboring islands exchange
individuals at some given rate.

It is often convenient to describe the state of the system in terms of
the macroscopic density of individuals $f(\mathbf{x},t)$ carrying one of the
two alleles. In the continuum limit, the macroscopic equation
governing the time evolution of such density reads
\begin{equation}\label{stochasticfisher}
  \partial_t f(\mathbf{x},t) = D \nabla^2 f(\mathbf{x},t)+ s f(1-f)
 +\sqrt{\frac{f(1-f)}{N}}\xi(\mathbf{x},t)
\end{equation}
where $N=N_l/a^d$, $a$ is the lattice spacing between two neighboring
islands\footnote{It is convenient to distinguish between $N_l$ (the
  population inside a single discrete deme of the
  SSM) and $N$ (the corresponding total density of individuals). The
  former is the quantity used to define the model, while the latter
  determines the amplitude of the noise due to number fluctuations in
  the continuum formulation of Eq. (\ref{stochasticfisher}). Notice
  that $N_l$ is a non-dimensional quantity, while $N$ is a density,
  carrying units of an inverse length to the power $d$ in $d$
  dimensions.}, $d$ the spatial dimension, and $\xi(\mathbf{x},t)$ is
a Gaussian stochastic process, delta correlated in space and time,
$\langle\xi(\mathbf{x},t)\xi(\mathbf{x}',t')\rangle=
\delta(\mathbf{x}-\mathbf{x}')\delta(t-t')$. Here, $f=1$ means an
island exclusively populated with one allele and $f=0$ means exclusive
occupation by the alternative genotype.  The nonlinearity multiplying
the noise requires an interpretation in terms of the Ito calculus
\citep{korolev_RMP}.

However, in many realistic cases, the mechanism of species movement
and range expansion is more complicated than a simple diffusion
process. For example, recent observations on crabs colonies along the
east coast of north America \citep{crabs} demonstrated how invasion of
one allele is controlled by the asymmetrical advection of larvae from
north to south by a coastal current. The interplay between population
genetics and individual movement (and transport) can be even more
complex in the open ocean, where individuals belonging to different
planktonic and bacterial species are stirred and mixed by chaotic
flows
\citep{tel_review,neufeld_book,dovidio,prasad_PRL,benzi_proc}. Of
particular interest is the population genetics of photosynthetic
organisms that control their buoyancy to remain near the surface of an
aquatic environment. In this case, the advecting flows are effectively
compressible, leading to population densities that overshoot the
normal carrying capacity \citep{prasad_PRL,pigo}.

While the SSM can be generalized to include a constant asymmetric
diffusion (see i.e. \citep{crabs}), the extension to more complex
fluid environments is more subtle. One of the main underlying
assumptions of the SSM -- a local population size that does not vary
either in time nor in space -- is quickly violated in aquatic
environments where flows create inhomogeneities in the total density
of individuals. Individual-based competition models without strict
population size conservation have already been studied, for example
allowing for the possibility of empty sites
\citep{neuhauser,omalley1,omalley2,omalley3,cencini}. However, when
flows are introduced, it is also less appropriate to discretize the
system in space into demes with a fixed size. In compressible
turbulence, for example, the density of individuals can be
inhomogeneous on a wide variety of spatial scales \citep{prasad_PRL},
even inside a single deme (which in the SSM is assumed to be
well-mixed).

In this paper, with the goal of describing population genetics in
aquatic environments in mind, we introduce a new model in which
individuals carrying two different alleles $A$ and $B$ live in a
continuous space. Their individual densities are allowed to grow and
fluctuate, including the important possibility of overshooting the
natural carrying capacity. Indeed, note that naively assuming
compressible flows that make $f>1$ would lead to an imaginary noise
amplitude in Eq. (\ref{stochasticfisher})!  The model we study is
similar in spirit to the stochastic logistic equation
\citep{law,lopez,birch}. However, in this study we focus on
competition and cooperation of {\em two} species, rather than the
stochastic growth of a single population. The second difference is
that previous studies focused on patterns formed by the non-local
nature of competition \citep{lopez,birch}. In this paper, we mostly
focus on the parameter range in which such patterns are not formed and
a weak noise description in the spirit of Eq. (\ref{stochasticfisher})
is appropriate.

The phenomenology of such a model, even in the presence of very simple
flows, is very rich due to the interplay between population dynamics
and fluid advection (see \citet{pigo} for some of the consequences in
one dimension). For this reason, we devote a large portion of this
work to the case in which the flow is absent and individuals move in
space in a diffusive way. This simple case allows for a systematic
comparison with the known results of the SSM. In particular, we show
that there exists a parameter range where the predictions of our model
are consistent with Eq. (\ref{stochasticfisher}) and its
generalization to include competitive exclusion and mutualism
\citep{korolev_mut}. In simple cases, such as when the two species are
neutral variants of each other, this correspondence can be shown
analytically. In more complex cases, the correspondence is explored by
means of numerical simulations. The last part of the work discusses an
example in which a compressible flow transport the individuals, as an
example of a problem that cannot be treated within the context of the SSM.

In Sec. \ref{mod_sec}, we sketch the model of growth, competition and
cooperation studied here, which leads to the two-species model for
allele densities $c_A(\mathbf{x},t)$ and $c_B(\mathbf{x},t)$
summarized in Eq. (\ref{eq_general}). We focus on three interesting
cases: (1) strictly neutral competitions, (2) a reproductive advantage
of one species over the other and (3) mutualistic situations where
cooperation plays a role. Sec. \ref{sec_mf} discusses the behavior of
our model in the ``zero-dimensional'' well-mixed case in which the
population is not structured in space, which allows us to determine
limits such that standard Wright-Fisher and Moran results for
population genetics can be recovered from our more general model.  We
then explore in Sec. \ref{section_space} the long-time behavior of our
model without fluid advection in one and two spatial
dimensions. Examples of the behavior of the model in the presence of
fluid advection are discussed in Sec. \ref{flowsection}. Concluding
remarks are presented in Sec. \ref{conclusions}. A detailed derivation
of our model equation is contained in Appendix A. Appendix B shows how
conventional stepping stone model results can be recovered in certain
limits. Appendix C describes a limit in which a mutualistic
generalization of the famous Kimura formula for fixation probabilities
\citep{crow} is possible.

\section{Model}\label{mod_sec}

Many widely studied models of population genetics in space, the most
notable example being the stepping stone model, consider individuals
carrying different alleles that occupy sites (also called ``demes'')
on a lattice. It is commonly assumed that each site is always
saturated up to its carrying capacity, so that, at each deme, the
local population size $N_l$ is constant during the dynamics.

We relax these assumptions by considering discrete individuals $X_i$
carrying different alleles (denoted by the index $i$) and diffusing in
continuous space (with a diffusion constant $D$, for simplicity equal
for all individuals). Further, we implement population dynamics
assuming that individuals carrying allele $i$ reproduce at rate
$\mu_i$ and die with rates $\tilde{\lambda}_{ij}$ proportional to the
number of individuals carrying a (possibly) different allele $j$ in a
region of spatial size $\delta$ centered on their position. For
example, in one dimension ($1d$), $\delta$ will be an interaction
length, while in $2d$ it will be an interaction area.  In a language
borrowed from chemical kinetics, the ``reactions'' we consider are:

\begin{eqnarray}\label{reactions}
X_i & \stackrel{\mu_i}{\rightarrow} & 2X_i\qquad 
\mathrm{(reproduction)}\nonumber\\
X_i + X_j & \stackrel{\tilde{\lambda}_{ij}}{\rightarrow} &
X_i\qquad \mathrm{(death}\ \mathrm{by}\ \mathrm{competition)}
\end{eqnarray}

In the case of a single species, this set of reactions is commonly
referred to as the birth-coagulation process \citep{doering_physA}.
In this paper, we will focus on the case of two alleles,
$i=A,B$. Other reactions could be added to the ones above, for example
the possibility that an individual can die even in absence of
competition, $X_i\rightarrow \emptyset$, or reactions
implementing more complex biological interactions. We will limit
ourselves to the biological dynamics embodied in (\ref{reactions}),
which contains minimal ingredients necessary to generate most of the
main features present in more complicated models. Notice that, in
contrast to models such as the Moran process, the density of
individuals is not fixed but fluctuates both locally and globally.

In order to make the presentation more compact, we start by discussing
the spatially explicit version of the model and then discuss the
globally well-mixed version as a limiting case. We consider the number
densities $n_A(\mathbf{x},t)$ and $n_B(\mathbf{x},t)$, that integrated
over a region of space yield the (stochastic) number of individuals of
species $A$ or $B$ in that region. We will study cases in which the
number densities are typically large, and consequently define
concentrations $c_A(\mathbf{x},t)=n_A(\mathbf{x},t)/N$ and
$c_B(\mathbf{x},t)=n_B(\mathbf{x},t)/N$ via a constant parameter $N$,
assumed to be of the same order of magnitude of $n_A$ and $n_B$. This
means that, by definition, a constant density $c=1$ corresponds to a
uniform distribution of $N$ individuals in a segment of length $1$ in
one dimension. More generally, in $d$ dimensions, a concentration
$c(\mathbf{x},t)=1$ will correspond to a total number of particles
$\mathcal{N}=NL^d$ in a system of linear size $L$.  
With this choice, the macroscopic equations describing
the dynamics of the concentrations $c_A$, $c_B$ of species $A$ and $B$
read:

\begin{eqnarray}\label{eq_general}
\frac{\partial}{\partial t} c_A & = & D\nabla^2 c_A + c_A( \mu_A - \lambda_{AA} c_A - \lambda_{AB} c_B) 
+\sqrt{\frac{c_A(\mu_A + \lambda_{AA} c_A + \lambda_{AB} c_B) }{N}}\xi \nonumber\\
\frac{\partial}{\partial t} c_B & = & D\nabla^2 c_B + c_B (\mu_B- \lambda_{BA} c_A - \lambda_{BB} c_B) 
+\sqrt{\frac{c_B (\mu_B+ \lambda_{BA} c_A + \lambda_{BB} c_B) }{N}}\xi'
\end{eqnarray}

where $\xi(\mathbf{x},t)$ and $\xi'(\mathbf{x},t)$ are independent
Gaussian random variables, delta-correlated in space and time,
$<\xi(\mathbf{x},t)\xi(\mathbf{x},t')>=\delta(t-t')\delta(\mathbf{x}-\mathbf{x}')$
that should be interpreted according to the Ito prescription
\citep{korolev_RMP}.  The macroscopic binary reaction rates
$\lambda_{ij}$ multiplying the quadratic terms in the concentrations
are defined in terms of the microscopic binary rates
$\tilde{\lambda}_{ij}$ as $\lambda_{ij}=N\delta\tilde{\lambda}_{ij}$,
where $\delta$ is the interaction domain defined above.  In the
following, we will focus on cases in which the $\mu_i$'s and the
$\lambda_{ij}$'s are of the same order of magnitude, so that typical
values of the total concentration $c_A+c_B$ are order $1$. Under these
assumptions, it is useful to note that the quantity
$2N^{-1}=2\delta/(\tilde{\lambda}_{ij}/\lambda_{ij})$ plays here the
same role of the genetic diffusion constant in the stepping stone
model. In particular, $\delta$ is analogous of the lattice spacing,
while the denominator on the right hand side can be thought as the
carrying capacity of each deme. A detailed derivation of
Eqs. (\ref{eq_general}), together with a discussion of its limits of
validity, is presented in Appendix A.  If the species densities are
well-mixed and we neglect stochastic number fluctuations, the
deterministic dynamics embodied in Eqs. (\ref{eq_general}) is a
familiar model of growth, selection and competition in asexual
populations \citep{maynardsmith}. The four different types of dynamics
that emerge depending on the values of the $\lambda_{ij}$'s are
reviewed at the end of this section. Our aim here is to understand the
rich behaviors possible when {\em both} spatial variations and number
fluctuations are allowed.

To limit the parameter space, we will consider the following
three biologically relevant choices for the reaction rates:

\begin{enumerate}
\item{Neutral Theory}

  This choice is appropriate when the two biological species (or
  strains, or mutants and wild type alleles) are neutral variants of
  each other. This means that their growth rates and carrying
  capacities are the same; further, competition with an individual
  belonging to the same species is the same as competition with an
  individual of the other species. In formulas, for
  Eq. (\ref{eq_general}), a convenient neutral parameter choice is:
  $\mu_A=\mu_B=\lambda_{AA}=\lambda_{AB}=\lambda_{BA}=\lambda_{BB}=\mu$.

\item{Reproductive advantage} 

  In this setting, we depart from neutrality by allowing for a
  different reproduction rate of species $A$: $\mu_A=\mu(1+s)$ while
  all the other rates (including the $\lambda_{ij}$) are equal to
  $\mu$ as in the neutral case.  We will study this case to explore
  the effect of a selective advantage of one of the two species on the
  dynamics of the model. In particular, $s>0$ implies a selective
  advantage for $A$ and $s<0$ is a disadvantage.  Clearly, neutrality
  is recovered for $s=0$.

\item{Mutualistic setting} 

  A simple way to study mutualistic interactions is to assume that the
  only departure from neutrality occurs in the intensity of
  competition between individuals carrying different alleles. In
  formulas, we have $\mu_A=\mu_B=\mu$, $\lambda_{AA}=\mu$,
  $\lambda_{BB}=\mu$, $\lambda_{AB}=\mu(1-\epsilon_A)$, and
  $\lambda_{BA}=\mu(1-\epsilon_B)$.  The corresponding macroscopic
  equations are well defined only for $\epsilon_A,\epsilon_B\le1$, so
  that the competition rates $\lambda_{ij}$ are non-negative. We will
  focus mostly on the case $\epsilon_A>0$ and $\epsilon_B>0$. In this
  regime, spatial number fluctuations play an important role
  \citep{korolev_mut} and competition between species is reduced (we
  will interpret this reduction as the effect of mutualistic
  interactions). Other choice could also be of interest, for example
  $\epsilon_A=0$ and $\epsilon_B<0$ is another way of allowing a
  competitive advantage of $A$ over $B$ (in this case, via enhanced
  competition rather than via a larger reproduction rate). We note
  finally that $\epsilon_A<0$, $\epsilon_B<0$ corresponds to a
  competitive exclusion model, arising for example when the competing
  variants secret toxins that inhibit the growth of their competitors.

\end{enumerate}

\noindent In the following, we will measure time in units of a
generation time so that $\mu=1$. A convenient choice of the
interaction domain is of the order of the average spacing among
individuals, $\delta=1/N$, so that
$\lambda_{ij}=\tilde{\lambda}_{ij}$. This choice also implies $N_l=1$.
For simplicity, we will present most of the spatial results for the
one-dimensional version of the model, introducing two-dimensional
results only as appropriate. In the spatially explicit case, the
system is a segment of length $L$ with periodic boundary
conditions. We will present also two dimensional simulations, where
the system is a $L\times L$ square, also with periodic boundary
conditions.

An even simpler setting we will study to make contact with traditional
Moran or Fisher-Wright models is the case in which the population can
be assumed to be well-mixed, or ``zero-dimensional''. This limiting
case can be easily obtained from the one dimensional case by setting
$\delta=L=1$ and ignoring spatial diffusion, since each individual now
interacts with every other individual in the population. As a
consequence of this choice, one now has
$N=\lambda_{ij}/\tilde{\lambda}_{ij}$. In this case, the spatial
position of the individuals is irrelevant for biological
interactions. Clearly, in this special case, the individual density is
equivalent to the total number of individuals $N \equiv N_l \equiv
\mathcal{N}$.

Both in the spatial and well-mixed cases, we will compare analytical
predictions obtained from the continuum theory of
Eqs. (\ref{eq_general}) with simulations of the individual-based
dynamics encoded in the reactions of Eqs. (\ref{reactions}). Details
on the numerical scheme implemented for the individual-based model are
in Appendix A and in \citep{prasad_proc}.

\begin{figure}[htb]
\begin{center}
\includegraphics[width=14cm]{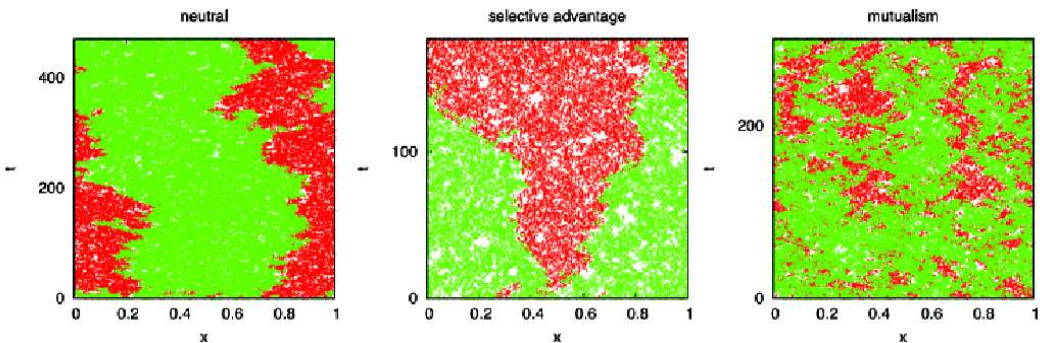}
\caption{Three illustrative parameters choices in the one
  dimensional version of the model. In all panels 
  $D=10^{-4}$ and $N=100$. The left panel corresponds to the neutral choice
  in which all rates are set to one and initially the two species are 
randomly distributed with equal concentrations. In the center panel,
 all parameters  are set to one except the reproduction rate of allele 
$A$ (in red) which 
reproduces at a rate $(1+s)$ with a large selective advantage $s=0.3$; 
in this case, the initial fraction of
 $A$ is $0.1$. In the right panel, competition 
among species is reduced by taking $\epsilon_A=\epsilon_B=0.7$ to enhance mutualism; in this 
case the two species are randomly distributed with equal concentrations in the initial condition. In this case, mutualism insures that the species (or alleles) remain spatially inhomogeneous out to very long times.
\label{fig_ill}}
\end{center}
\end{figure}

In Fig. (\ref{fig_ill}), we anticipate some of the results to
illustrate the qualitative behaviors that can be explored with the
three aforementioned parameter choices in one spatial dimension. In
the left panel, the two alleles are neutral. Despite fluctuation of
the total density, the phenomenology is similar to that of the $1d$
stepping stone model: as time progresses, the two alleles are demixed
and fixation occurs by coalescence of the domain boundaries, which can
be regarded as annihilating random walks. In the central panel,
species $A$ (in red) initially constitutes only $10\%$ of the total
population; however, it has a reproductive advantage over species
$B$. Despite the discreteness of individuals and density fluctuations,
there are two noisy Fisher waves by which the initial minority can
take over the entire population. Finally, in the right panel we
simulate a case in which mixing of the two species is promoted by
reducing competition among different alleles. In this case, we expect
the two species to remain mixed indefinitely in the limit of large
system size.

In the remainder of this section, we introduce some of the concepts we
want to investigate in the simple case of a well mixed system without
number fluctuations.  Intuition about mutualistic behavior (and its
opposite, competitive exclusion \citep{frey}) can be obtained by
neglecting both the spatial degrees of freedom and the noise terms in
Eq. (\ref{eq_general}). In this simple case, the dynamics reduces to
\citep{korolev2012}
\begin{eqnarray}\label{eq_LV}
\frac{d}{dt}n_A(t)=n_A(t)\left[\mu_A-\tilde{\lambda}_{AA}n_A(t)-\tilde{\lambda}_{AB}n_B(t)\right]\nonumber\\
\frac{d}{dt}n_B(t)=n_B(t)\left[\mu_B-\tilde{\lambda}_{BA}n_A(t)-\tilde{\lambda}_{BB}n_B(t)\right].
\end{eqnarray}
Note that the intrinsic carrying capacities (i.e., the steady state
densities of one species when the other is absent) for this model are
$N_A=\mu_A/\tilde{\lambda}_{AA}$ and $\qquad
N_B=\mu_B/\tilde{\lambda}_{BB}$.  These quantities (we always choose
parameters such that $N_A\approx N_B$) play the role of the parameter
$N$ that controls stochastic number fluctuations in the general case
of Eq. (\ref{eq_general}).  As mentioned above for case 3, an
especially interesting situation arises when (1) the two species grow
at identical rates when the numbers are dilute, so that
$\mu_A=\mu_B=\mu$; (2) also the self-competition terms are also
identical, $\tilde{\lambda}_{AA}=\tilde{\lambda}_{BB}$; and (3) the
effect of cooperation or competitive exclusion is contained
exclusively in the cross-interactions,
$\tilde{\lambda}_{AB}\equiv\tilde{\lambda}_{AA}(1-\epsilon_A)$ and
$\lambda_{BA}\equiv\tilde{\lambda}_{BB}(1-\epsilon_B)$.  With this
choice, and rescaling the time unit by a factor $\mu^{-1}$, the
equations for the concentrations $c_A=n_A/N_A$ and $c_B=n_B/n_B$
corresponding to system (\ref{eq_LV}) read
\begin{eqnarray}\label{mut_LV}
\frac{d}{dt}c_A = c_A \left[1-c_A-c_B+\epsilon_A c_B\right]\nonumber\\
\frac{d}{dt}c_B = c_B \left[1-c_A-c_B+\epsilon_B c_A\right].
\end{eqnarray}

\begin{figure}[htb]
\begin{center}
\includegraphics[width=13cm]{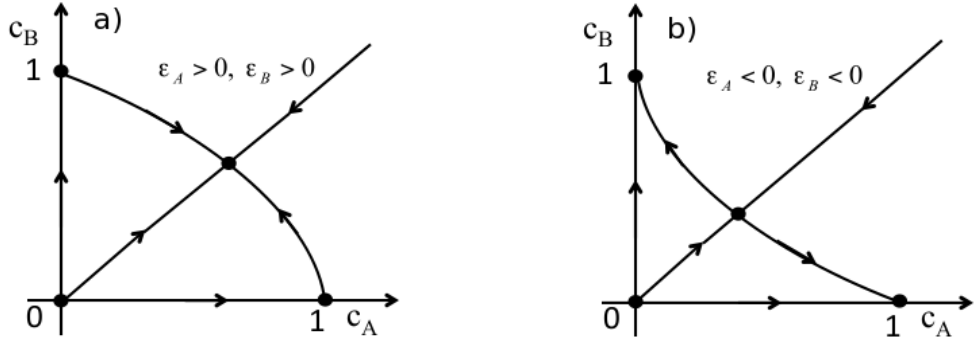}
\caption{Deterministic dynamics of the mutualistic model in zero
  dimensions without number fluctuations.  In a), the interactions
  $\epsilon_A>0$ and $\epsilon_B>0$ favor cooperation, and there is a
  stable fixed point $(c_A^*,c_B^*)$ with both densities nonzero. In
  B), the organisms secrete toxins that impede each others growth, so
  $\epsilon_A<0$ and $\epsilon_B<0$ and the fixed point
  $(c_A^*,c_B^*)$ is unstable.
\label{drnfig}}
\end{center}
\end{figure}

The remaining two parameters $\epsilon_A$ and $\epsilon_B$ control the
competition under ``crowded conditions'', such that the populations
have grown up to satisfy $c_A+c_B\approx 1$. If the two variants are
nearly identical, it is reasonable to assume $|\epsilon_A|$,
$|\epsilon_B|\ll 1$.  As illustrated in Fig. \ref{drnfig}, the
deterministic system (\ref{mut_LV}) always has fixed points at
$(0,0)$, $(0,1)$, and $(1,0)$. Depending on the parameters, there can
also be a fourth fixed point \citep{maynardsmith} located at
\begin{equation}\label{fixp}
  (c_A^*,c_B^*)=\frac{(\epsilon_A,\epsilon_B)}{\epsilon_A+\epsilon_B-\epsilon_A\epsilon_B}.
\end{equation}
When cooperation is favored ($\epsilon_A,\epsilon_B>0$, Fig. \ref{drnfig}a) this
fixed point is stable, and leads to a steady state population fraction
$f^*$ of $A$ individuals, $0< f^*<1$, with
\begin{equation}
f^*\equiv \frac{c_A^*}{c_A^*+c_B^*}=\frac{\epsilon_A}{\epsilon_A+\epsilon_B}.
\end{equation}
When competitive exclusion \citep{frey} is favored
($\epsilon_A,\epsilon_B<0$, fig. \ref{drnfig}b) this fixed point is
unstable to the attracting fixed points $(1,0)$ or $(0,1)$, depending
on the initial conditions. Genetic demixing, present in strictly
neutral systems only due to stochastic number fluctuations, is {\em
  enhanced} in this case. Finally, when $\epsilon_A$ and $\epsilon_B$
have {\em opposite} signs, the fixed point (\ref{fixp}) lies outside
the biologically relevant domain, and one of the two fixed points
$(1,0)$ or $(0,1)$ becomes globally stable, corresponding to a
competitive advantage for one species or the other when the population
is dense.

Suppose we now introduce spatial migration and number fluctuations, to
recover the full model defined by Eq. (\ref{eq_general}). When might
we expect fixation probabilities, the global heterozygosity,
correlation functions etc. to reduce to the familiar results for
conventional spatial stepping stone-type models with strictly
conserved population sizes in every deme? A particularly simple case,
corresponding to the selectively neutral limit
$\epsilon_A=\epsilon_B=0$, is illustrated for a well-mixed system in
Fig. 3a below: the population grows up and eventually wanders along
the line $c_A+c_B=1$, until it reaches the absorbing states at $(1,0)$
or $(0,1)$. A more general situation is $\epsilon_A+\epsilon_B=0$, in
which case one variant typically has a simple selective advantage
along an invariant subspace given by the line $c_A+c_B=1$. If the
fluctuations transverse to this line are small (corresponding to a
large population size), then the usual formulas for fixation
probabilities hold, as we show later in this paper. In more general
situations, however, it is no longer exactly true that the population
localizes at long times near the straight line $c_A+c_B=1$. Indeed, we
have from Eq. (\ref{fixp}) that
\begin{equation}\label{fixp2}
c_A^*+c_B^*=\frac{\epsilon_A+\epsilon_B}
{\epsilon_A+\epsilon_B-\epsilon_A\epsilon_B},
\end{equation}
which exceeds 1 along the outwardly bowed incoming trajectories in
Fig. \ref{drnfig}a, and is less than $1$ for the outgoing inwardly
curved trajectory in Fig. \ref{drnfig}b. However, we do have the
approximate equality, $c^*_A+c^*_B\approx 1$, provided
$|\epsilon_A+\epsilon_B|\ll|\epsilon_A\epsilon_B|$ in
Eq. (\ref{fixp2}). In this limit, a combination of numerical and
analytic arguments presented in this paper show that formulas recently
derived for mutualistic and competitive exclusion stepping stone
models \citep{korolev_mut} apply to the current model with demographic
fluctuations as well, again provided that the overall population size
$N$ is sufficiently large.

What happens if $\mu_A$ and $\mu_B$ are unequal, but $\epsilon_A$ and
$\epsilon_B$ remain small?  In this case, the population proportions
will certainly change as an initially small population like that in
Fig. 3a grows to approach the line $c_A + c_B \approx 1$.  However,
once this line is reached, the subsequent time evolution should again
be given by stepping stone model results.

\section{Well-mixed case with number fluctuations} \label{sec_mf}

In this section, we present the results in the simple well-mixed (or
``zero-dimensional'') version of the model. Thus, we keep number
fluctuations in Eq. (\ref{eq_general}), but neglect spatial variations
in the allele concentrations.

\subsection{Neutral theory}

As previously discussed, it is useful to describe the dynamics of the
neutral version of the model in the $c_A$ vs. $c_B$ plane, as depicted
in Fig. (\ref{neutral_MF_fig}, left). Starting from a dilute initial
condition, the system evolves rapidly towards to the intrinsic overall
carrying capacity given by $c_A+c_B=1$. The dynamics is then localized
near this line (with fluctuations), until one of the two species goes
extinct. This behavior contrasts with the Moran process in which the
dynamics is rigidly confined to the $c_A+c_B=1$ line, since no
fluctuations of the total density are allowed.  To determine when these
fluctuations are small, first note from
Eq. (\ref{eq_general}) that in the neutral case the total
concentration $c_T=c_A+c_B$ obeys a closed equation:

\begin{figure}[htb]
\begin{center}
\includegraphics[width=12cm]{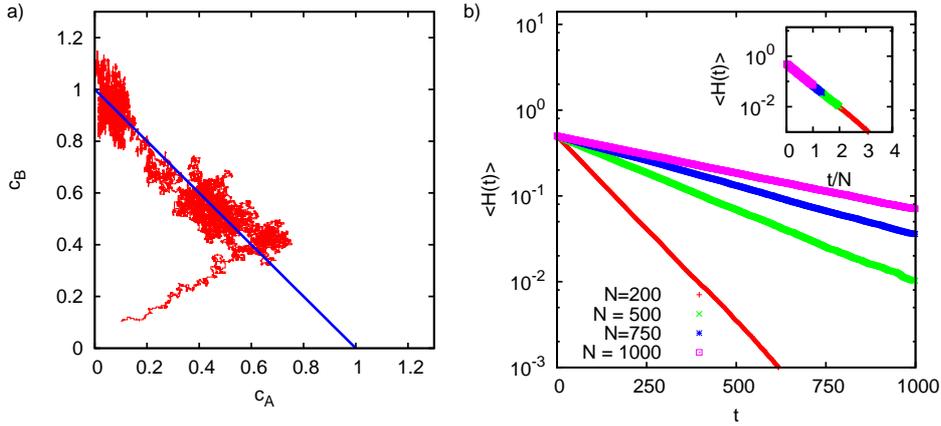}
\caption{Neutral dynamics in the well-mixed case.
(a) Example of a trajectory in the $(c_A, c_B)$ plane with
  $N=500$. The initial condition is $n_A=n_B=20$, i.e. a small
  fraction of a typical long time carrying capacity.  (b) Decay of
  the average heterozygosity $\langle H(t)\rangle$ for different
  values of $N$. Curves are obtained from simulations of the particle
  model; each curve is an average over $10^4$ realizations and
  the error bars are smaller than the size of the lines. (inset) Same
  curves plotted as a function of $t/N$. Note the data collapse.
  \label{neutral_MF_fig}}
\end{center}
\end{figure}

\begin{equation}\label{eq_ct}
\frac{d}{dt}c_T =\mu c_T(1-c_T) + \sqrt{\frac{\mu c_T(1+c_T)}{N}}\xi_c,
\end{equation}
decoupled from the fraction of species $A$, $f=c_A/(c_A+c_B)$, where the noise
term $\xi_c$ satisfies $\langle\xi_c(t)\xi_c(t')\rangle=\delta(t-t')$.
When $N$ is large, the stationary solution, beside the solution
$P(c)=\delta(c)$ corresponding to global extinction that will
eventually be reached \footnote{Notice that, as in the particle model
  for simplicity death is implemented only via binary reactions (see
  Eq. \ref{reactions}), the state of global extinction is not
  accessible in the particle model. Such discrepancy with the
  macroscopic equation could be easily removed by allowing for death
  even in absence of competition, i.e. the reaction $X_i\rightarrow
  \emptyset$.} on long times of order $\exp(N)$, is approximately a
Gaussian with average $\langle c_T\rangle=1$ and variance $\langle
c_T^2\rangle-\langle c_T\rangle^2=N^{-1}$, which is small when $N$ is
large.

We now describe the dynamics of the relative fraction $f$ of individuals
carrying allele $A$, $f=c_A/(c_A+c_B)$. The equation for $f(t)$, derived in
 \ref{app_f}, reads

\begin{equation}\label{neutr_eq}
\frac{d}{dt}f = \sqrt{\mu f(1-f)\frac{1+c_T}{Nc_T}}\xi_f.
\end{equation}

where $\xi_f(t)$ also satisfies
$\langle\xi_f(t)\xi_f(t')\rangle=\delta(t-t')$, and further we have
$\langle\xi_f(t)\xi_c(t')\rangle=0$.  The above equation allows us to analyze the
global heterozygosity, which quantifies the loss of diversity as time
evolves and is defined as the probability $H(t)=2\langle
f(1-f)\rangle$ that two randomly chosen individuals in the population
carry different alleles.

As mentioned above, the equation for $c_T$ is independent of $f$ in
the neutral case studied here.  As a result, one can factorize
the average over $c_T$ and $f$ in the equation for $H(t)$:

\begin{equation}
\frac{d}{dt} H(t)=-\frac{\mu}{N}\left\langle f(1-f)\frac{1+c_T}{c_T}
\right\rangle=-\frac{\mu}{N}\left\langle f(1-f)\right\rangle
\left\langle\frac{1+c_T}{c_T}\right\rangle=
-\frac{2\mu}{N}H(t)+O\left(\frac{1}{N^2}\right).
\end{equation}

Neglecting the correction of order $N^{-2}$, we recover for our model
with density fluctuations the closed equation for $H(t)$ for
Fisher-Wright and Moran-type models with a fixed population size
derived by Kimura, which states that the total heterozygosity decays
exponentially in well mixed neutral systems \citep{crow}:

\begin{equation}
\langle H(t)\rangle=H(0)\exp(-2\mu t/N)
\end{equation}

Fig. (\ref{neutral_MF_fig}b) confirms this exponential behavior in
simulations of the model.

\subsection{Reproductive advantage}

In a well-mixed finite population and in absence of mutations,
diversity will be lost and only one of the two alleles will survive
after a long enough time.  We now study the probability of allele $A$
to fixate in a well-mixed population of size $N\gg1$, in the case in
which the allele confers a small reproductive advantage $s\ll1$.  In
the same spirit as the previous section, we can derive the equation
for the relative fraction $f=c_A/(c_A+c_B)$ (see \ref{app_f}).  Upon
neglecting terms proportional to $s/N$, the equation in this case
reads:

\begin{equation}\label{adv_eq}
\frac{d}{dt}f =\mu s f(1-f) +\sqrt{\mu f(1-f)\frac{1+c_T}{Nc_T}}\xi
\end{equation}

As in Eq. \ref{neutr_eq}, this result must be supplemented with the
equation for the total concentration $c_T=c_A+c_B$. Although in the
non-neutral case the equation for $c_T$ is no longer independent of
$f$, one can show that the averages over $c_T$ and $f$ factorize up to
terms of order $s/N$ or higher that can be safely neglected for $s\ll
1$ and $N\gg 1$.

\begin{figure}[htb]
\begin{center}
\includegraphics[width=13cm]{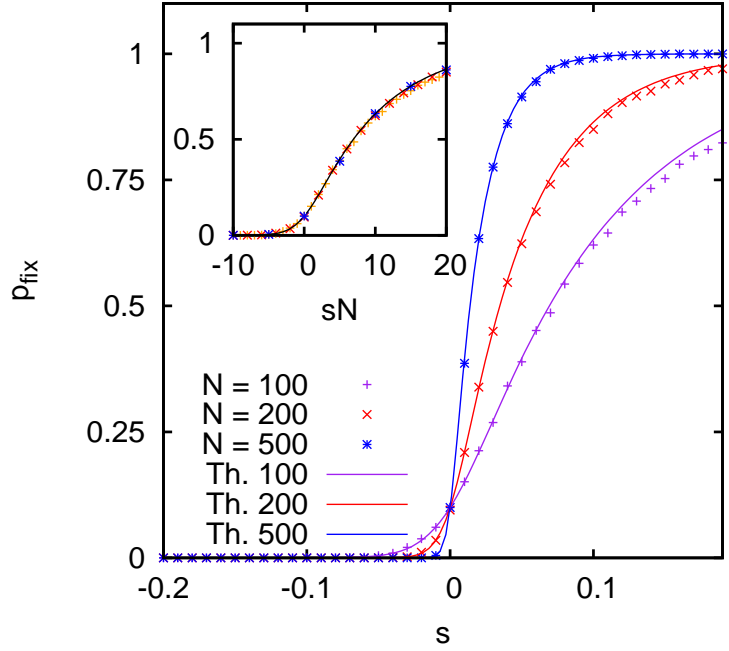}
\caption{Fixation in the well-mixed case with reproductive advantage. 
Probability of fixation for different values of $s$ and $N$ 
(listed in the figure) for the well-mixed version of the particle model,
  compared with the prediction of Eq. (\ref{kimurafix}). The initial 
fraction of individuals belonging to species $A$ is $f_0=0.1$, with $c_T=1$ initially. 
The inset  shows that all curves collapse when plotted as a function 
of $sN$. These curves are again averages over $10^4$ independent 
realizations.
\label{fig_select_mf}}
\end{center}
\end{figure}

These observations allows us to recover the formula for the probability
of fixation of allele $A$ \citep{crow}.

\begin{equation}\label{kimurafix}
p_{fix}=\frac{1-e^{-sNf_{0}}}{1-e^{-sN}}
\end{equation}
where $f_0$ is the fraction of individuals carrying allele $A$ once
trajectories like that in Fig (\ref{neutral_MF_fig}a) reach the line
$c_A+c_B=1$.  This result is again similar to Fisher-Wright or Moran
models with a strictly fixed total population
size. Eq. (\ref{kimurafix}) is tested with simulations in
Fig. \ref{fig_select_mf}.

\subsection{Mutualism}

In the well-mixed limit of the mutualistic model, fixation always
occurs at $(c_A,c_B)=(1,0)$ or $(0,1)$ after a long enough
time. However, when the total number of individuals is large, this
time grows exponentially with $N$ and can easily become inaccessible
to experiments (and simulations). As detailed in 
\ref{app_mut_mf}, the quasi-stationary solution where the two
cooperating species coexist for $\epsilon_A,\epsilon_B>0$ can be seen
as a state confined by two potential barriers, one inhibiting species
$A$ to fixate and the other inhibiting species $B$ to fixate. When $N$
is large, it will be extremely probable that fixation will occur by
passing the lowest of these two barriers. In this case, an estimate of
the time $t^*$ needed to reach fixation can be derived by calculating
the height of the lowest barrier and applying Kramer's escape rate
theory. The result is:

\begin{equation}\label{kramer}
t^*\sim\exp\left[\frac{N}{2}\frac{\min(\epsilon_A^2,\epsilon_B^2)}
{\epsilon_A+\epsilon_B}\right].
\end{equation}

\begin{figure}[htb]
\begin{center}
\includegraphics[width=12cm]{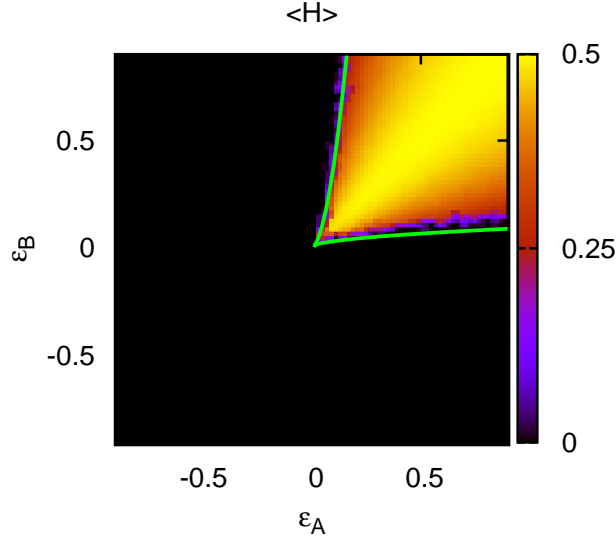}
\caption{Finite-time coexistence in the well-mixed mutualistic model.
  Average heterozygosity in the $(\epsilon_A, \epsilon_B)$ plane, with
  $N=500$, in $d=0$ dimensions, i.e. for the well-mixed model.
  Simulations are run until a time $t=5000$. For each pair of
  $(\epsilon_A, \epsilon_B)$ values, after a transient, the
  heterozygosity approaches a quasi-stationary value.  The green line
  limits the region in which coexistence up to this time is possible
  according to the estimate (\ref{kramer}).\label{het_mut_fig}}
\end{center}
\end{figure}

Figure (\ref{het_mut_fig}) shows a heat map of the total
heterozygosity in the ($\epsilon_A, \epsilon_B$ plane for $N=500$
after $5000$ generations. The black region is where fixation
occurred. Green lines are the theoretical limits of the apparent coexistence
region obtained from Eq. \ref{kramer}.

After estimating the fixation time in the mutualistic model, we now
ask: what is the fixation probability of one of the two alleles? In
Appendix C, we show that in the appropriate limit the fixation
probability for mutualists obeys a formula similar to the result for a
stepping stone model with {\em fixed} total population size
\citep{korolev_mut}, namely

\begin{equation}
u(f_0)=\frac{\int_0^{f_0}e^{\frac{1}{2}Ns(f^*-p)^2}}{\int_0^{1}e^{\frac{1}{2}Ns(f^*-p)^2} },
\end{equation}
where $f_0$ is the initial fraction of allele $A$. In the limit
$f^*\rightarrow\infty$, $s\rightarrow 0$, with a mutualistic effective
selective advantage $\tilde s=f^* s$ fixed, this reduces to the famous
Kimura formula discussed above

\begin{equation}
u(f_0) ~=~ \frac{1-e^{-N \tilde s f_0}}{1-e^{-N \tilde s}}
\label{Kimur}
\end{equation}

The formulas above are a good approximation for arbitrary initial
conditions only for the case of equal initial growth rates
$\mu_A=\mu_B=\mu$, so that population fractions are approximately
unchanged prior to reaching the line $c_A+c_B\approx 1$.  We
explore the fixation probabilities in three different cases, each
having a different definition of selective advantage:

\begin{figure}[htb]
\includegraphics[width=0.8\textwidth]{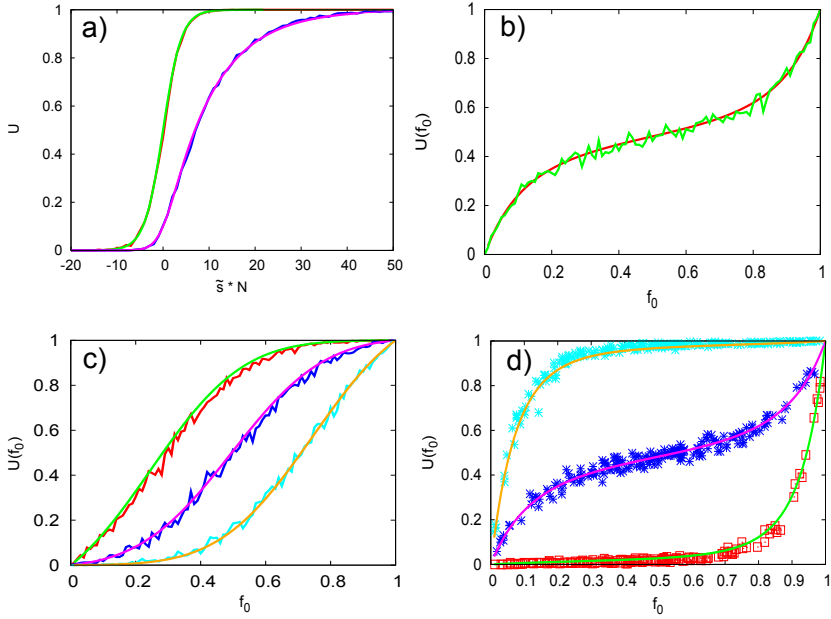}
\caption{Fixation probabilities $u(\tilde{s},f_0,N)$in the mutualistic
  model.  Full curves show the analytical results from
  Eq. (\ref{fixation}) with initial fraction $f_0 =
  n_A(0)/(n_A(0)+n_B(0)$.  a) Competitive exclusion: Simulations with
  $\epsilon_A + \epsilon_B=0$ with $-0.08 < \epsilon_A < 0.2,N=250$:
  Red curve: $f_0=0.1$, blue curve $f_0=0.5$. Green and purple curves:
  Eq. (\ref{Kimur}) with $-0.08 < \tilde s < 0.2,N=250$, $f_0=0.1$
  (green), $f_0=0.5$ (purple).  The curves are plotted against the
  scaling variable $\tilde s * N$ for different initial frequencies
  $f_0$. Here (and also in b),c)) the initial condition is chosen on
  the line $n_A(0)+n_B(0)=N$.  b) Mutualism: Green curve: simulations
  with $\epsilon_A= \epsilon_B =0.1, N=100, f^*=0.5$. The fixation
  probability $u$ is plotted versus the initial fraction $f_0$.  Red
  curve: Fixation formula (\ref{Kimur}) with $N=100,\tilde
  s=0.1,f^*=0.5$.  c) Coordination game with an unstable fixed point
  $f^*$: Green; purple; orange curves: simulations with
  $\epsilon_A=-0.05,\epsilon_B=-0.15 (f^*=0.25);
  \epsilon_A=-0.10,\epsilon_B=-0.10
  (f^*=0.50);\epsilon_A=-0.15,\epsilon_B=-0.05 (f^*=0.75)$ .  Red;
  blue; cyan curves: Fixation formula (\ref{fixation}) with
  $N=100,\tilde s/f^*=-0.2,f^*=0.25;0.5;0.75$.  d) Mutualism with
  stochastic initial conditions.  Simulations with initial conditions
  $n_A(0),n_B(0)$ are uniformly distributed in the plane of size $N \times N$
  with $N=100$.  For each random initial condition, which fixes the
  value of $f_0$, the fixation probability is averaged over 500
  independent Gillespie simulations resulting in $u(f_0)$.  Cyan
  points: $\epsilon_A=0.05,\epsilon_B=0.15,f^*=0.25$; blue points:
  $\epsilon_A=0.10,\epsilon_B=0.10,f^*=0.5$; red points:
  $\epsilon_A=0.15,\epsilon_B=0.05f^*=0.75$.  Full curves: fixation
  formula (\ref{fixation}) with $N=100,\tilde s/f^*=0.2$: Brown:
  $f^*=0.25$; purple: $f^*=0.25$; green $f^*=0.75$.}
\label{zerod}
\end{figure}

\begin{itemize}
\item $\epsilon_A + \epsilon_B = 0$. Unless $\epsilon_A=\epsilon_B=0$,
  this corresponds to a selective advantage under crowded conditions,
  such that $c_A+c_B\approx 1$. In the previous section, we discussed
  how in the deterministic limit there are two stable fixed points,
  $(c_A^*,c_B^*)=(1,0)$ and $(c_A^*,c_B^*)=(0,1)$, while the fixed
  point with both $c_A^*$ and $c_A^*$ nonzero is inaccessible.
  Fig. \ref{zerod}a shows the fixation probability for
  $c_A(t=0)+c_B(t=0)=1$ and two initial frequencies $f_0=0.5$,
  $f_0=0.1$, $N = 250$, $f^*=\epsilon_A/(\epsilon_A+\epsilon_B)$ and
  effective selective advantage $\tilde s=\mu
  \epsilon_A=-\mu\epsilon_B$.  The population size $N$ appears through
  the combination $\tilde s * N$ in Eq. \ref{fixation}, so we plot the
  probability versus this rescaled parameter.  We obtain  excellent
  agreement between this special case of our model and the Kimura
  formula for the Moran model Eq. (\ref{Kimur}).

\item $\epsilon_A + \epsilon_B = \tilde s/f^*,\epsilon_A > 0,
  \epsilon_B > 0$. This corresponds to a mutualistic situation in which
  there is a stable fixed point out in the plane $(c_A^* > 0,c_B^* >
  0)$. Fig. \ref{zerod}b shows the fixation probability $u(f_0)$
  versus the initial fraction $f_0$ for stochastic Gillespie
  simulations with $\epsilon_A = \epsilon_B = 0.1$ where $f^*=0.5$ and
  $N=100$. For comparison, the formula Eq.(\ref{fixation}) is shown as
  the full drawn line again indicating very good agreement.

\item $\epsilon_A + \epsilon_B = -\tilde s/f^*, \epsilon_A < 0,
  \epsilon_B < 0$.  This choice corresponds to the competitive
  exclusion \citep{frey} in which there is an unstable fixed point in
  the plane $(c_A^* > 0,c_B^* > 0)$ and two stable fixed points where
  one of the two species has gone extinct. Fig. \ref{zerod}c shows
  Gillespie simulations for three cases of $\epsilon_A <0,~ \epsilon_B
  < 0$ and a comparison with the formula Eq. (\ref{fixation}) for the
  different values of $f^*$ (in order to compare this case we take
  $\tilde s < 0$ in the formula).
 
\end{itemize}

As a further case we consider random initial conditions $n_A(0),
n_B(0)$ uniformly distributed in the square $[1,N]\times[1,N]$, so
that the approach to the line $c_A+c_B\approx 1$ can play a role as
well. The initial fraction is now defined as $f_0 =
n_A(0)/[n_A(0)+n_B(0)]$. Fig. \ref{zerod}d show the corresponding
Gillespie simulation results for 200 different initial conditions for
three different fractions $f^* = 0.25,0.5,0.75$. The analytic fixation
curves according to Eq. (\ref{fixation}) are also shown. Although the
agreement is excellent, we again expect modification when departures
from equality of the initial growth rates $\mu_A$ and $\mu_B$ are
allowed.

\section{One and two dimensions} \label{section_space}

\begin{figure}[htb]
\begin{center}
\includegraphics[width=13.5cm]{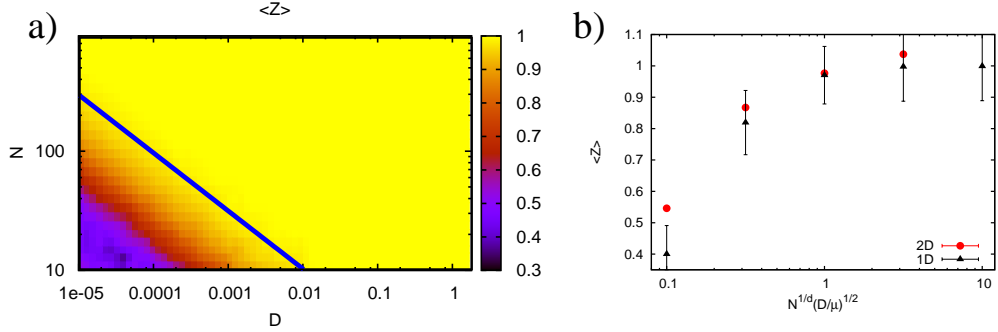}
\caption{Reduction of the carrying capacity in the neutral model 
in $1d$ and $2d$.
(a) Reduction of the total carrying capacity 
$Z=\langle c_A+c_B\rangle$
in the $(D, N)$-plane. The system is one dimensional 
and we adopted the neutral choice of parameters (see Section 2). 
The blue line is the theoretical condition $N\sqrt{D}/\mu =1$
. (b) Comparison of the carrying capacity reduction in $1d$ and $2d$, 
as a function of the nondimensional parameter $N^{1/d}\sqrt{D/\mu}$ where 
$d$ is the spatial dimension.
\label{carrying_fig}}
\end{center}
\end{figure}

Density fluctuations play a more significant role in one and two
spatial dimensions, compared with the well-mixed situations described
in the previous section. For example, depending on initial conditions
and genetic drift, different alleles can fix in different regions of
space; the ultimate fate of the system then depends on how these
different regions interact, which in turn depends on the choice of the
rates. One of the most striking effects of spatial variation in allele
number and relative proportions is the existence of a regime in which
there is a reduction in the average carrying capacity, i.e. the
average concentration $Z$ is smaller than the value $\langle Z \rangle
= 1$ calculated from Eqs. (\ref{eq_general}) with our choice of
parameters and by neglecting fluctuations. The presence of such a
regime is illustrated in Fig. (\ref{carrying_fig}) in the neutral case
as a function of the $D$ and $N$. Notice that the latter parameter can
be properly interpreted as an average number of particles per unit
length only when $N$ and $D$ are both large enough. In the opposite
regime, as a consequence of fluctuations, the average number of
particles in a unit segment is significantly less than $N$. We
quantify this effect by defining an effective average carrying
capacity $\langle Z \rangle=\langle n(t)\rangle/N$ where $n(t)$ is the
actual number of particles present at time $t$ per unit length and the
average $\langle\dots\rangle$ is over time.

We find significant deviations from the prediction $\langle Z\rangle
=1$ when $N\sqrt{D/\mu}\ll1$ . Heuristically, this criterion can be
understood as follows. In spatially extended systems, the populations
are mixed by diffusion. The diffusion scale $\sqrt{D/\mu}$ may be
considered as an ``effective deme size'', in the sense that
individuals within a distance less than $\sqrt{D/\mu}$ are mixed very
efficiently over a single generation, while individuals separated by a
larger distance are spatially decoupled. In one dimension, the
condition $N\sqrt{D/\mu}\gg1$ then corresponds to having many
individuals in an effective deme size. In the opposite limit, this
number is small and fluctuations play a much more important role. This
effect is related to the ``strong noise limit'' of the stochastic
Fisher equation (see e.g. \citet{doering_physA,berti,korolev_str}). We
remark that in this regime, the assumptions needed to derive
Eq.(\ref{eq_general}) from the particle model are violated and
significant deviations between the particle simulations and the
macroscopic theory are expected. For this reason, we will restrict our
analysis here to the ``weak noise'' case in which $N\sqrt{D/\mu}>1$.

\subsection{Neutral theory}

To study how fixation occurs in space, we now discuss the behavior of
the spatial heterozygosity $H(x,t)$ defined as the probability of two
individuals at distance $x$ and time $t$ to carry different
alleles. In the neutral stepping stone model with a fixed population
size in each deme, $H(x,t)$ obeys a closed equation:

\begin{equation}\label{ev_H}
\partial_t H(x,t)=D\nabla^2 H-\frac{2\mu}{N}H \delta(x).
\end{equation}

\noindent In one dimension, such equation can be solved explicitly:

\begin{equation}\label{het1d_th}
H(x,t)=H_0\left[1-\frac{2}{N}\int_0^t\ dt' 
\frac{\mathrm{erf}\left(\frac{t'}{4N^2D}\right)}{\sqrt{8\pi D(t-t')}}
e^{-\frac{x^2}{8D(t-t')}+\frac{t'}{4N^2D}}\right]
\end{equation}

\noindent where $H_0$ is the initial heterozygosity, equal to one half
if the two variants are well mixed and equally populated at time
$t=0$. Eqs. (\ref{ev_H}) and (\ref{het1d_th}) can be derived directly
from the stochastic Fisher equation (\ref{stochasticfisher}) with
$s=0$ (see, e.g., \citet{korolev_RMP}).

We define the heterozygosity in our off-lattice particle simulations
with growth and competition from the statistics of interparticle
distances. In particular, at a given time $t$, we compute all 
distances between pairs of individals. Upon introducing a bin size
$h$, the function $H(r,t)$ is then defined as the ratio between the
number of pairs carrying {\em different} alleles at a separation
between $r$ and $r+h$, divided by the total number of pairs of all
types in the same range of separation. For simplicity, we always took
the bin size $h$ equal to the interaction distance $\delta$.

\begin{figure}[htb]
\begin{center}
\includegraphics[width=12cm]{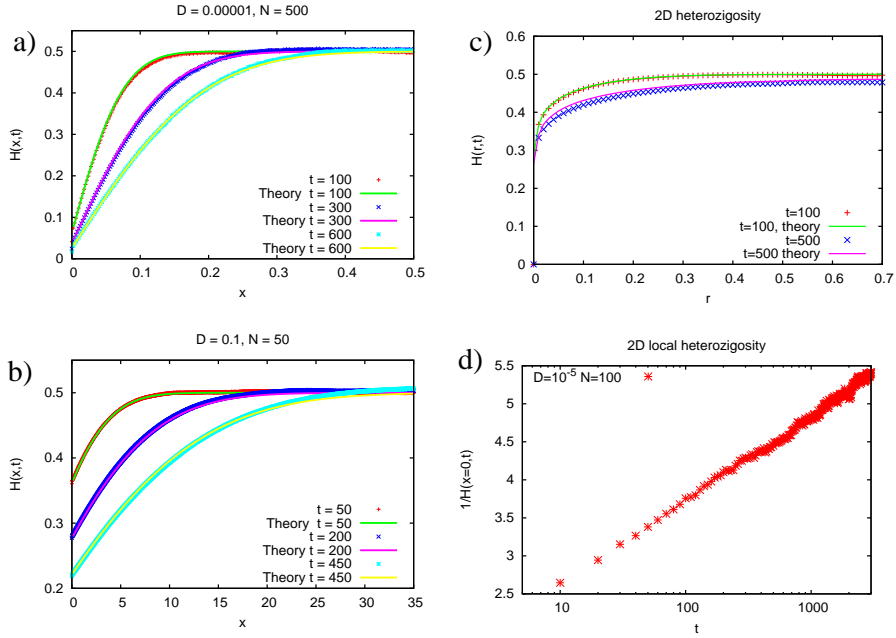}
\caption{Heterozygosity in the $1d$ and $2d$ neutral case.
Behavior of heterozygosity correlation function for the neutral 
off-lattice model of growth and competition. (a) 1D simulations
  at low diffusivity, $D=10^{-5}$ and (b) high diffusivity,
  $D=0.1$.  In the top case, the system size is $L=1$ while in the
  bottom case the system size is $L=100$. In both cases we find
  excellent agreement with the prediction of formula
  (\ref{het1d_th}). (c) Neutral heterozygosity in $2d$, compared
  with a numerical integration of Eq. (\ref{ev_H}). (d)
  Behavior of the local heterozygosity $H(x=0,t)$ as a function of
  time in 2D, showing the logarithmic decay $H(x=0,t)\sim 1/\ln(t)$.
  \label{het1d_fig}}
\end{center}
\end{figure}

In the limit $N\sqrt{D/\mu}\gg1$, the spatial heterozygosity obtained
by simulations of the neutral off-lattice model shows a remarkable
agreement with Eq. (\ref{het1d_th}), as shown in
Fig. (\ref{het1d_fig}). This correspondence arises because the
relative fraction of allele $A$, $f(x,t)=c_A/(c_A+c_B)$, obeys a very
similar equations as discussed in the mean field case. In Appendix B,
we show that the only effect of density fluctuation is an additional
effective advection term in the equation for $\partial_t f$, equal to
$2D(\nabla \log c_T)\cdot \nabla f$. The appearance of such term was
already found in \citet{Vlad2004} in a deterministic version of the
model described here.  In our case, one can show that since $c_T$
obeys a decoupled equation in the neutral case, such term will not
affect the equation for the heterozygosity. Indeed, numerical
simulation shows that the average spatial heterozygosity in the model
reproduces that of the stepping stone model even in the limit of very
high diffusivity shown in Fig. \ref{het1d_fig}, panel (b). Panel (c)
shows that similar agreements arise comparing numerical integration of
Eq. (\ref{ev_H}) with our off-lattice simulations in two
dimensions. At variance with one dimension, where the local
heterozygosity $H(0,t)$ decays at long times as $t^{-1/2}$, in two
dimension the decay is much slower, $H(0,t)\sim 1/\ln(t)$. Such slow
logarithmic decay is confirmed in simulations in panel (d).

\subsection{Reproductive advantage}

\begin{figure}[htb]
\begin{center}
\includegraphics[width=12cm]{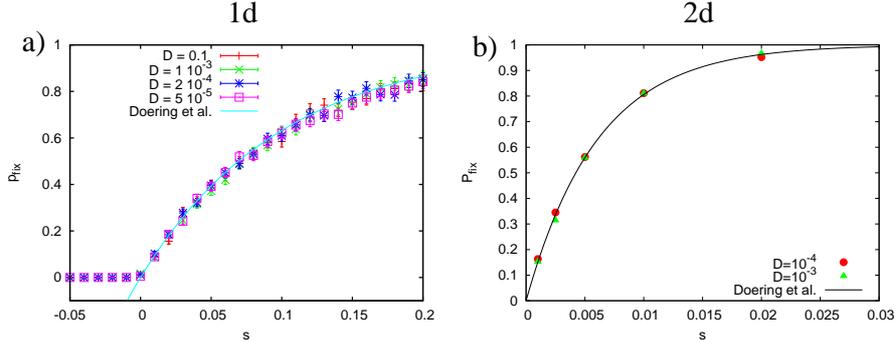}
\caption{Probabilities of fixation in the presence of a reproductive advantage.
The two panels show (a) one spatial dimension and (b)
  two dimensions, as a function of the selective advantage $s$, 
for different values 
of the diffusion constant
  $D$. Our $1d$ results are compared the results with the
  prediction of \citet{doering_physA}. In 1d, parameters are $N=500$
  and the initial fraction of species A is $f_0=0.01$, randomly
distributed on the unit interval. The $2d$ simulations 
  were conducted on a square domain of unit area and the parameters $N=16384$ 
  and the initial fraction of species A is $f_0=0.01$ were kept fixed. 
  The solid line is our conjectured generalization of Eq.
(\ref{doer_form}) to two dimensions.
\label{fig_select_1d}}
\end{center}
\end{figure}

In one spatial dimension, an analogue of Kimura's formula
(\ref{kimurafix}) \citep{crow} for the fixation probability has been
derived from the stochastic Fisher equation by
\citet{doering_physA}:
\begin{equation}\label{doer_form}
p_{fix}=1-\exp\left[-sN\int\ dx \  f(x,t=0)\right]
\end{equation}
where $f(x,t=0)$ is the initial spatial distribution of the fraction
of species $A$.  Remarkably, the one dimensional fixation probability
is independent of the spatial diffusion 
constant. We tested this prediction in Fig. (\ref{fig_select_1d}a),
left panel, for our model when species $A$ enjoys a reproductive
advantage $s$. There are again no appreciable differences between the
simulations of our more general growth model and the theoretical
prediction for the stepping stone model, over a wide range of
diffusion constants. This agreement is expected, given the approximate
mapping onto a stepping stone model embodied in Eq. (B.3) of Appendix
B.  While the result (\ref{doer_form}) by \citet{doering_physA} was
derived in one dimension, we conjectured that the same formula holds
in two dimensions. Indeed, a straightforward generalization of
Eq. (\ref{doer_form}) predicts well the fixation probability in two
dimensions, as shown in simulations in Fig. (\ref{fig_select_1d}),
right panel.

\subsection{Mutualism}

\begin{figure}[htb]
\begin{center}
\includegraphics[width=13.5cm]{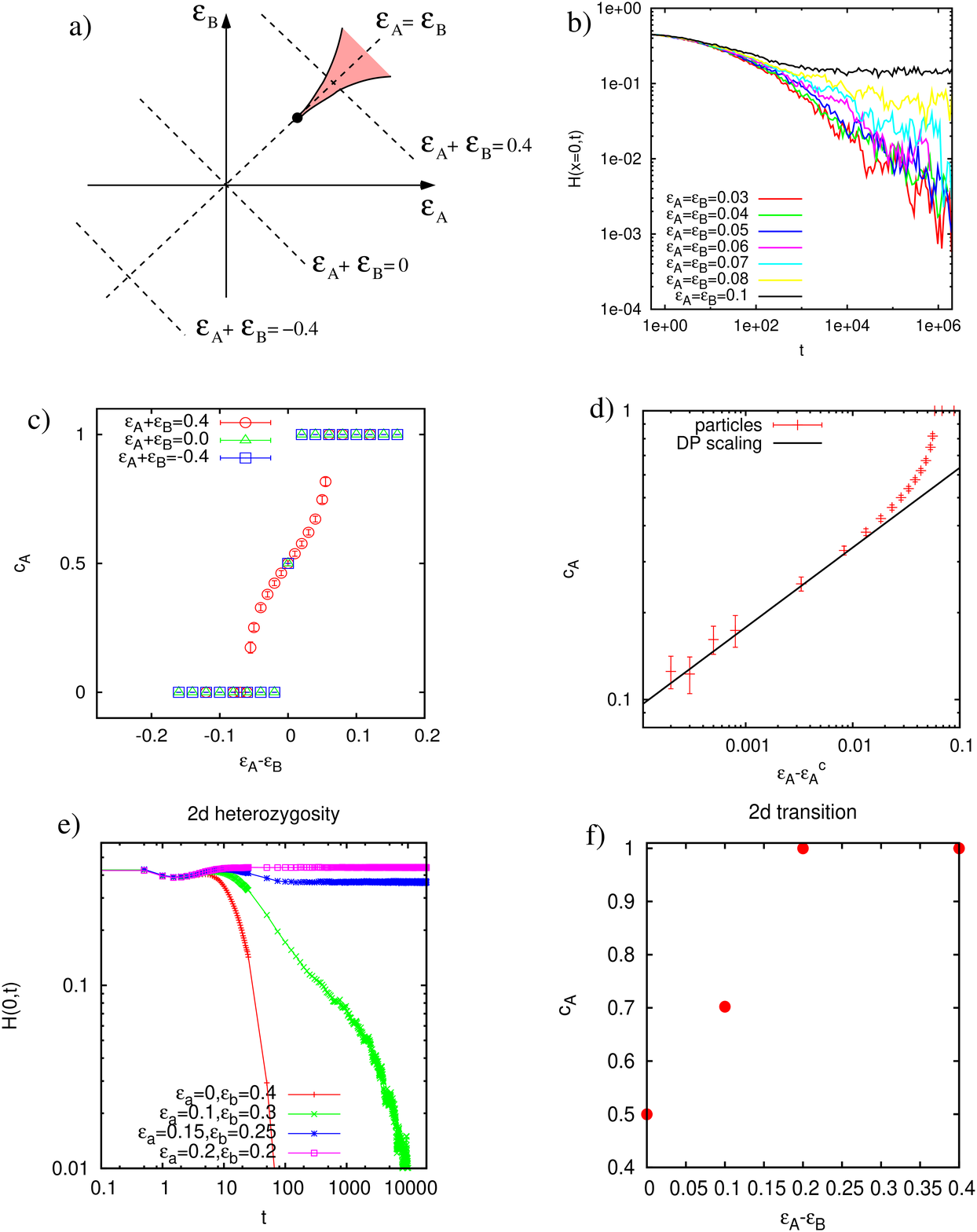}
\caption{Mutualism in $1d$ and $2d$.
a) Phase diagram of the mutualistic model in 1d. The
  mutualistic region, where global fixation never occurs in an
  infinite system, is colored in red. Dashed lines denote the cuts
  relevant to data in the other panels. b) Behavior of the local
  heterozygosity $H(0,t)$ along the cut $\epsilon_A=\epsilon_B$. A
  nonzero long time asymptote implies that fixation never occurs.  c)
  Average concentration of allele $A$, $<c_A>$, along three cuts such
  that $\epsilon_A+\epsilon_B=const.$. When $\epsilon_A+\epsilon_B$ is
  sufficiently large and positive, $\langle f \rangle$ varies smoothly
  between $0$ and $1$ when traversing the red region in (a). For both
  $\epsilon_A+\epsilon_B=0$ and $\epsilon_A+\epsilon_B$ negative,
  there is an abrupt jump in $\langle f\rangle$ from $0$ to $1$ when
  $\epsilon_A=\epsilon_B$.  In this sense, the dashed diagonal line
  below the cusp in (a) is like a first order phase transition.  In
  all figures, parameters are: $\mu=1$, $D=0.02$, $N=30$ and $L=2000$
  so that on average there are $6\cdot 10^4$ individuals in the
  system. d) Logarithmic plot of the density of $A$ close to the
  critical point. A power law with the expected directed percolation
  exponent, $f(x)\propto x^\beta$, $\beta\approx 0.2765$ is shown for
  comparison. e) Behavior of the local heterozygosity $H(0,t)$ in $2d$
  along the line $\epsilon_A=\epsilon_B$. A phenomenology similar to
  the $1d$ case of panel b) is observed.  f) Transition along the diagonal cut
  $\epsilon_A+\epsilon_B=0.4$ in $2d$, again showing a similar
  behavior to the $1d$ case shown in panel c).\label{phasemut} }
\end{center}
\end{figure}

We now set $\mu_A=\mu_B=\mu$, but allow variable interspecific
competition coefficients can vary in one and two dimensions.
\citet{korolev_mut} recently demonstrated how for a mutualistic
stepping stone model with fixed deme size in one dimension, there is a
region in the $(\epsilon_A, \epsilon_B)$ parameter space in which (in
limit of an infinite system size, $L\rightarrow \infty$) fixation
never occurs, as sketched in Fig. \ref{phasemut}, panel a). This
behavior differs dramatically from the well-mixed zero dimensional
case, for which fixation is inevitable, with a fixation time
$t^*(\epsilon_A,\epsilon_B,N)$ given approximately by
Eq. (\ref{kramer}).

We fix parameters as $\mu=1$, $D=0.02$ and $N=30$. To explore the
behavior of our model, we performed simulations along the paths shown
as dashed lines in panel a) of Fig. (\ref{phasemut}). Panel b) shows
the time evolution of the local heterozygosity $H(0,t)$ along the line
$\epsilon_A=\epsilon_B$. For small values of
$\epsilon_A=\epsilon_B>0$, the heterozygosity decays in a similar
fashion (roughly as $1/\sqrt{t}$) as in the neutral case
$\epsilon_A=\epsilon_B=0$. For higher values, the local
heterozygosity eventually levels off at a nonzero value, implying that
fixation will never occur.

The presence of a mutualistic regime where the system remains mixed
forever is even more evident in Fig. \ref{phasemut}, panel c), where
we plot along the cuts at constant $\epsilon_A+\epsilon_B$ the
long-time average of the fraction of the first allele $\langle
f\rangle$ as a function of the difference
$\epsilon_A-\epsilon_B$. Along the cuts that do not cross the
mutualistic region, $\langle f\rangle$ is either $0$ or $1$ as one of
the two alleles always fixes. A special case arises for
$\epsilon_A=\epsilon_B$, where each of the two alleles has a chance of
being fixated equal to its relative abundance in the initial
condition, so that $\langle f\rangle=f_0$. Conversely, $\langle f
\rangle$ has a non-trivial behavior along the line
$\epsilon_A+\epsilon_B=0.4$. Upon varying the parameter
$\rho=\epsilon_A-\epsilon_B$, we find a whole range of values in which
fixation does not occur. As discussed in \citep{korolev_mut}, the two
lines of critical points shown in (a) are in the directed percolation
universality class.  The behavior of the density close to this
critical point is described by a universal exponent,
$c_A\sim(\epsilon_A-\epsilon_c)^\beta$, where the expected exponent is
$\beta\approx 0.2765$ and $\epsilon_c$ is the value of $\epsilon_A$ at
the critical point (see e.g. \citet{odor}). Fig. \ref{phasemut}, panel
d) confirms the power law behavior close to one of the critical points
on the cut $\epsilon_A+\epsilon_B=0.4$. Finally, in panels e) and f)
we show simulations on the two dimensional mutualistic
model. Mutualism in $2d$ is computationally challenging and, to the
best of our knowledge, has not been studied systematically in the
literature. Although we did not obtain the full phase diagram, our
simulations suggest a scenario similar to the $1d$ case. In
particular, the heterozygosity $H(x=0,t)$ along the cut
$\epsilon_A=\epsilon_B$ displays a transition from a regime in which
it seems to decay logarithmically (as in the $2d$ neutral version of
the model) to a regime in which fixation does not occur. Furthermore,
the cut at $\epsilon_A+\epsilon_B=0.4$ shown in panel f) reveals a
directed-percolation-like transition, qualitatively similar 
to that in panel c).

\section{Population genetics in two-dimensional compressible turbulence}\label{flowsection}
A systematic exploration of the effect of hydrodynamic flows on the
off-lattice models of population genetics introduced here would take
us far beyond the scope of this already lengthy paper. However, to
illustrate the interesting effects that arise, we now extend our
analysis to the two cases where the competition between populations
takes place under the influence of compressible fluid advection in
two-dimensions. As we will show, compressible fluid flows can
dramatically change the carrying capacities and fixation times. For
all the simulations in this section we choose a square simulation
domain of size $[0,L]\times[0,L]$, the spatial diffusivity
$D=10^{-4}$.  For simplicity, the two competing populations are
neutral with
$\mu_A=\mu_B=\lambda_{AA}=\lambda_{AB}=\lambda_{BA}=\lambda_{BB}=1$.

The two flows that we choose are:
\begin{enumerate}
\item {\it Compressible surface flow (CSF)}: \\This chaotic,
  time-dependent flow is generated from a two-dimensional slice of a
  three-dimensional, homogeneous, isotropic flow (see
  \cite{prasad_PRL,prasad_aXv}). A snapshot of the advecting velocity
  field is shown on the left side of Fig.~\ref{pot_flow}. Using the
  projection method described in \cite{prasad_aXv} we choose the
  compressibility of the flow $\kappa=1$ where, $\kappa\equiv\langle
  (\nabla \cdot {\bf u})^2 \rangle/\langle (\nabla {\bf u})^2
  \rangle$, ${\bf u}\equiv(u_x,u_y)$ is the velocity field, and
  $\langle (\cdot) \rangle$ indicate the spatio-temporal
  averaging. Setting $\kappa$ to its maximum value of unity maximizes
  the reduction in carrying capacity caused by locally compressing the
  populations to high density, so that the middle terms on the right
  side of Eqn.~(3) are negative [\cite{pigo,prasad_aXv}]. The strength
  of the flow is varied by scaling the velocity field by a forcing
  amplitude $F$. For all the simulations with this flow we choose
  $L=2\pi$.

\item {\it Steady flow (SF)}: \\This time-independent velocity field
  is chosen to be $u_x(x,y)=F[ \alpha \sin(2\pi x/L) + (1-\alpha)
  \sin(2\pi y/L)]$, $u_y(x,y)=F [\alpha \sin(2\pi y/L) - (1-\alpha)
  \sin(2\pi x/L)]$ (see Fig.~\ref{pot_flow} right panel).  The 
  strength of the flow is controlled by again changing $F$ and the
  compressibility $\kappa=\alpha^2/[\alpha^2 + (1-\alpha^2)]$ is
  modified by changing $\alpha$ and hence $\kappa \in [0,1]$. For all
  the simulations with this flow we choose $L=1$.  The two species are
  advected by the flow towards the sink which is located at the center
  $(-L/2,L/2)$ of the simulation domain.
\begin{figure}[!h]
\begin{center}
\includegraphics[width=0.48\linewidth]{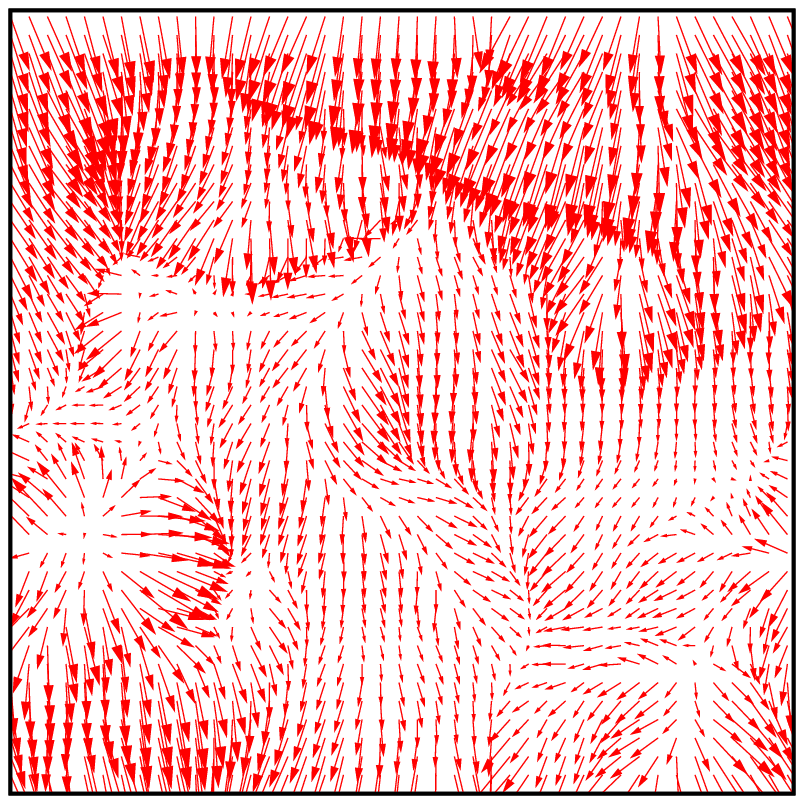}
\includegraphics[width=0.48\linewidth]{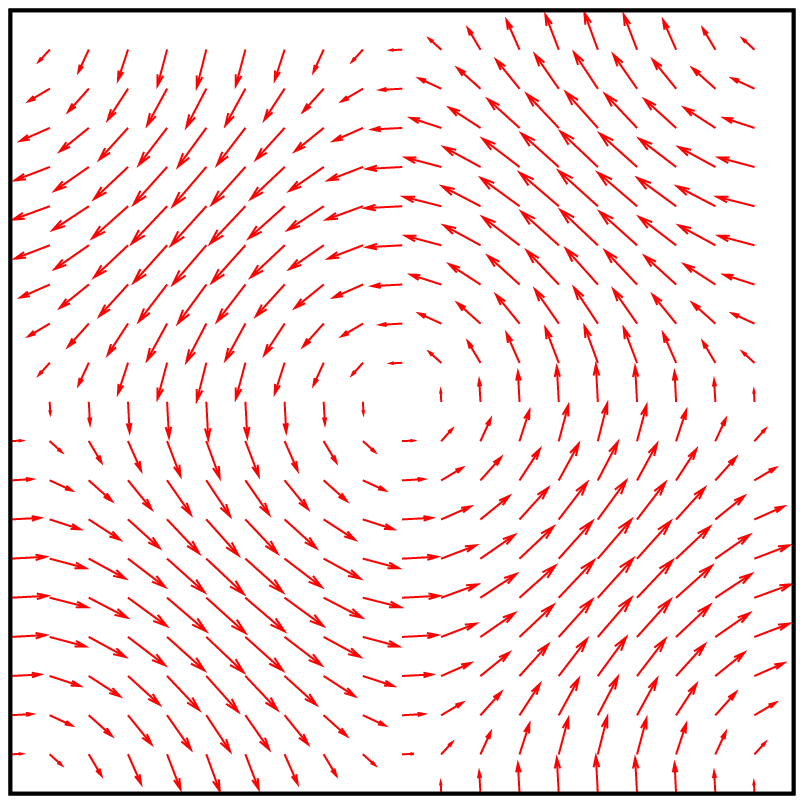}
\end{center}
\caption{\label{pot_flow} (Left) A representative snapshot of the
  time-dependent compressible surface flow (CSF) field used for
  advecting species in our two-dimensional simulations. (Right) Vector
  field visualization of the steady flow (SF) used for advecting
  species in our simulations of a simple time-independent steady flow
  with $\kappa=0.0027$.}
\end{figure}
\end{enumerate}
Similar analysis for one-dimensional flows was conducted in
\cite{pigo}. The compressible flow on the left of Fig.~\ref{pot_flow}
models photosynthetic organisms that control their buoyancy to remain
near the surface of a turbulent ocean. The flow on the right is
designed to determine the consequences for population genetics of
fluid sink at the center, with fluid injection at the four
corners. Note the non-zero vorticity in this case.

\begin{figure}[!h]
\begin{center}
\includegraphics[width=0.32\linewidth]{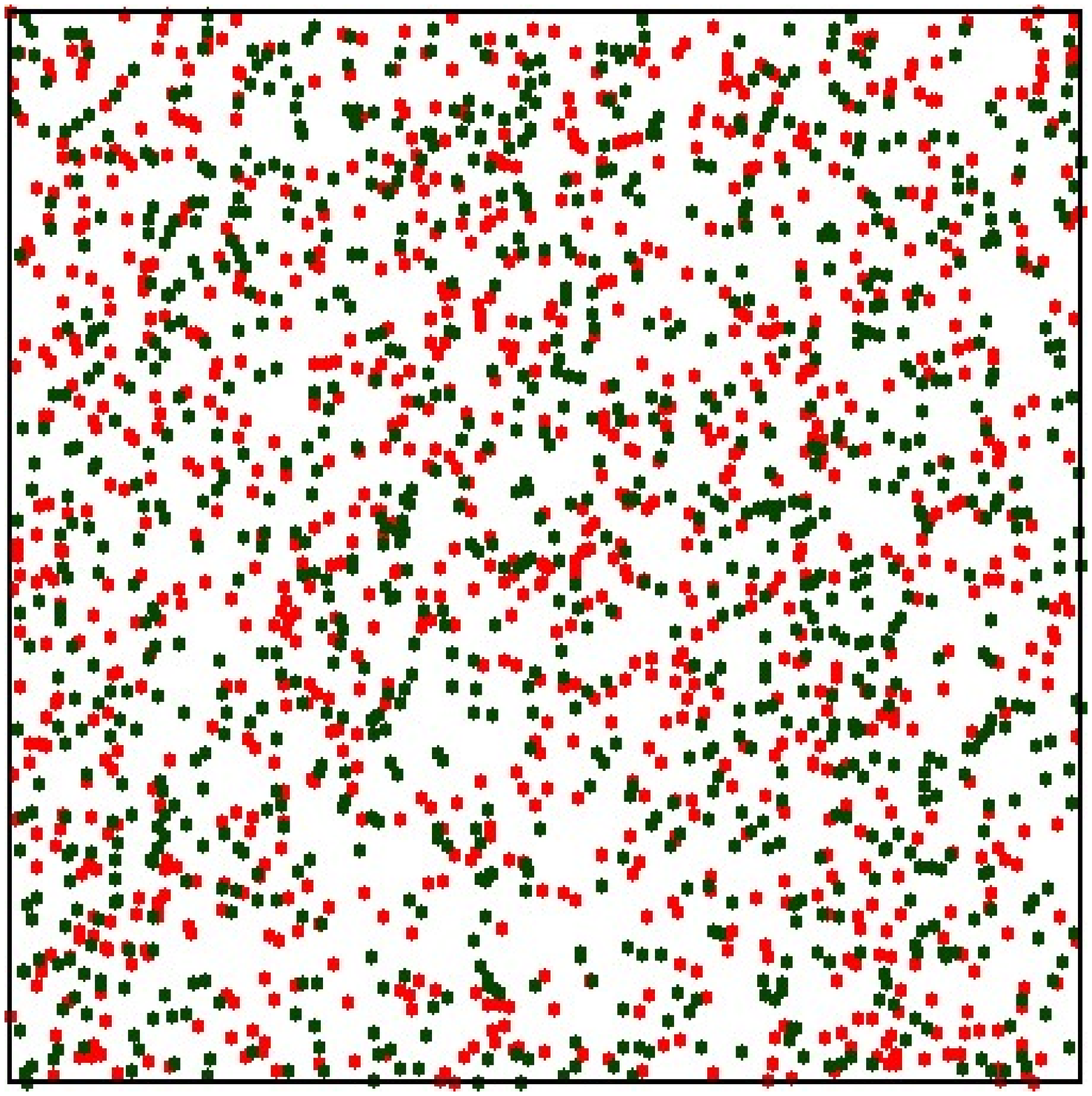}
\includegraphics[width=0.32\linewidth]{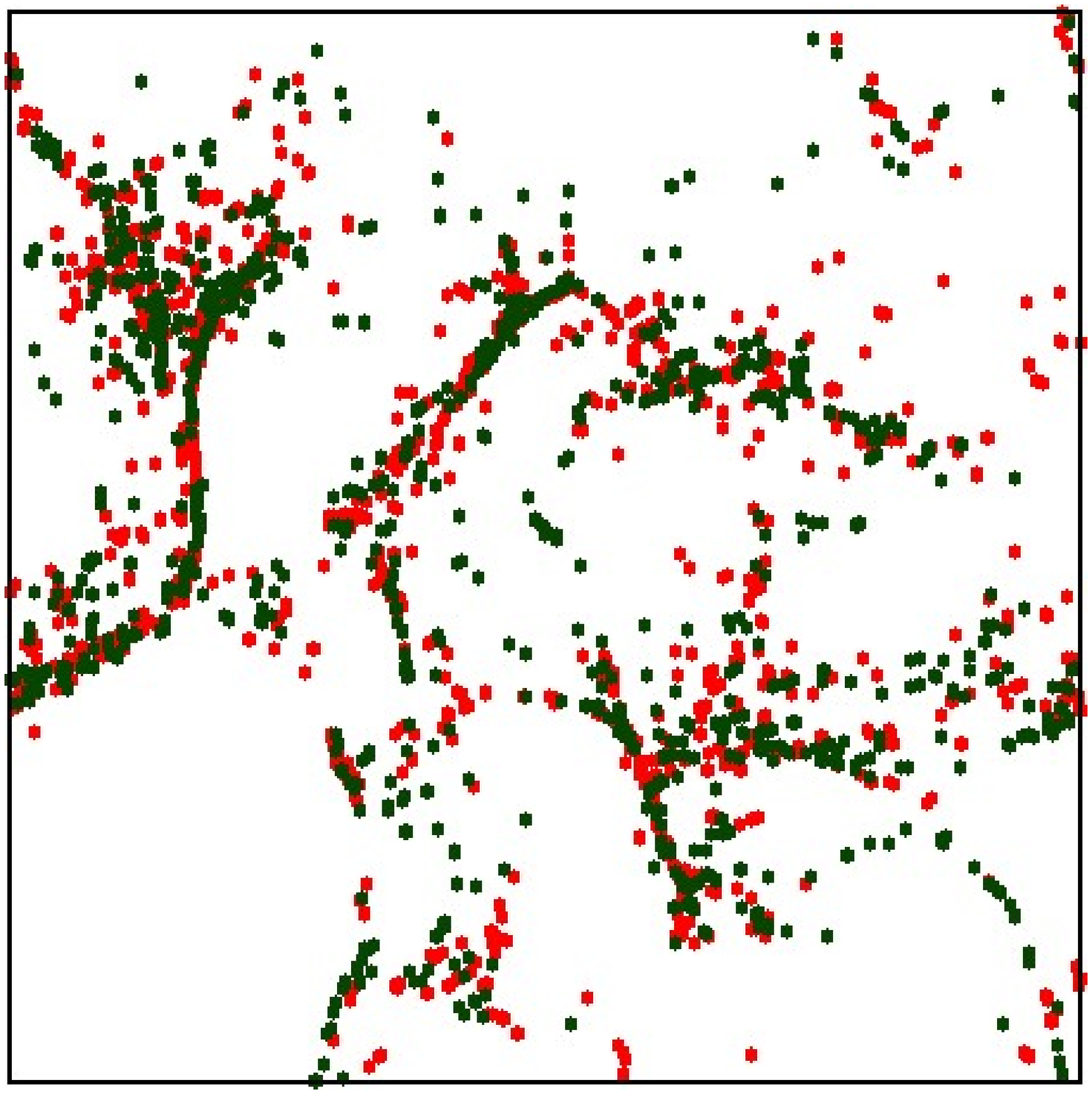}
\includegraphics[width=0.32\linewidth]{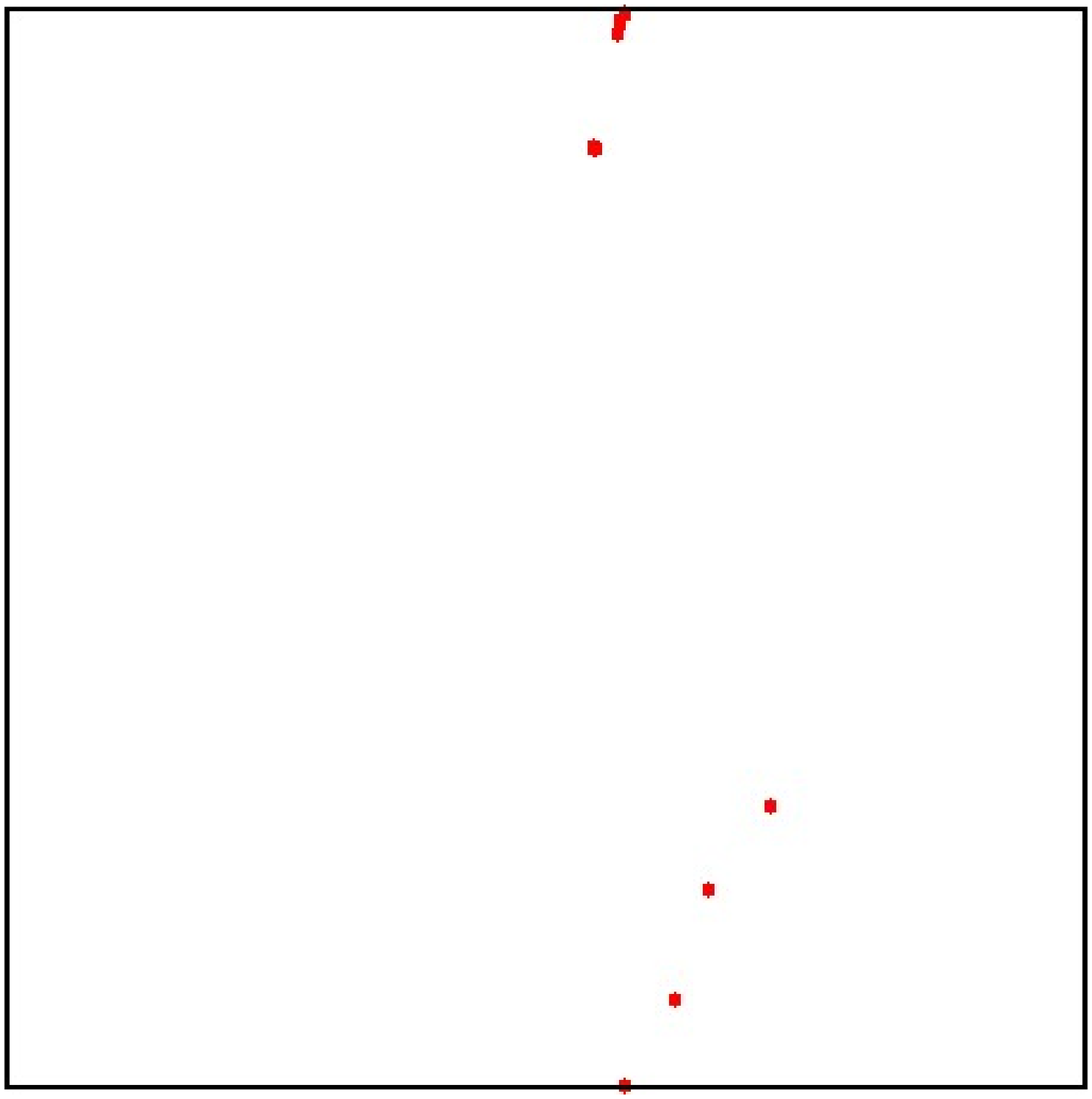}
\end{center}
\caption{\label{fig:csnap} Competition between two neutral species
  (shown in red and green) in a turbulent compressible flow with
  $\kappa=1$ and $F=1$. At time $t=0$ (left) approximately $10000$
  organisms are randomly distributed over the entire domain at the
  steady state carrying capacity in absence of flow. Both species are
  then collapsed by advection onto filamentous structures leading to
  (time-dependent) sinks and saddle points, dynamics which
  compactifies the population into regions where competition takes
  place. This collapse is highlighted in the middle plot which is
  chosen at a later time $t=1$ (middle). At much later times $t=25$
  (right) fixation occurs and only one of the species survive. The
  populations size has stabilized at $6$ organisms, a reduction from
  the initial carrying capacity by a factor $10^3$. Although the
  reduction in the population size is most extreme for $\kappa=1$,
  significant reductions occur for even small values of $\kappa$
  [\cite{prasad_aXv}].}
\end{figure}

\begin{figure}[!h]
\begin{center}
\includegraphics[width=0.32\linewidth]{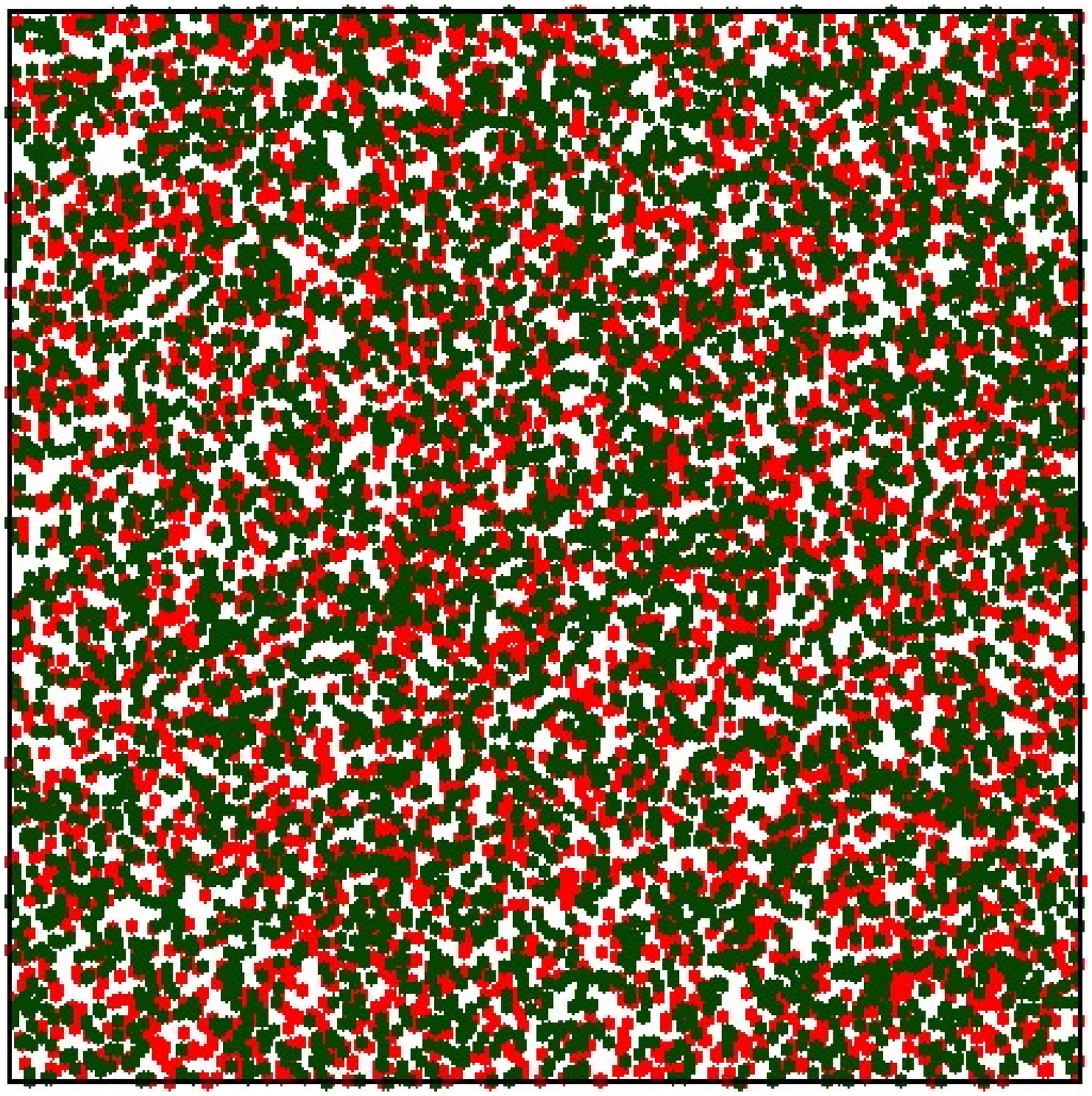}
\includegraphics[width=0.32\linewidth]{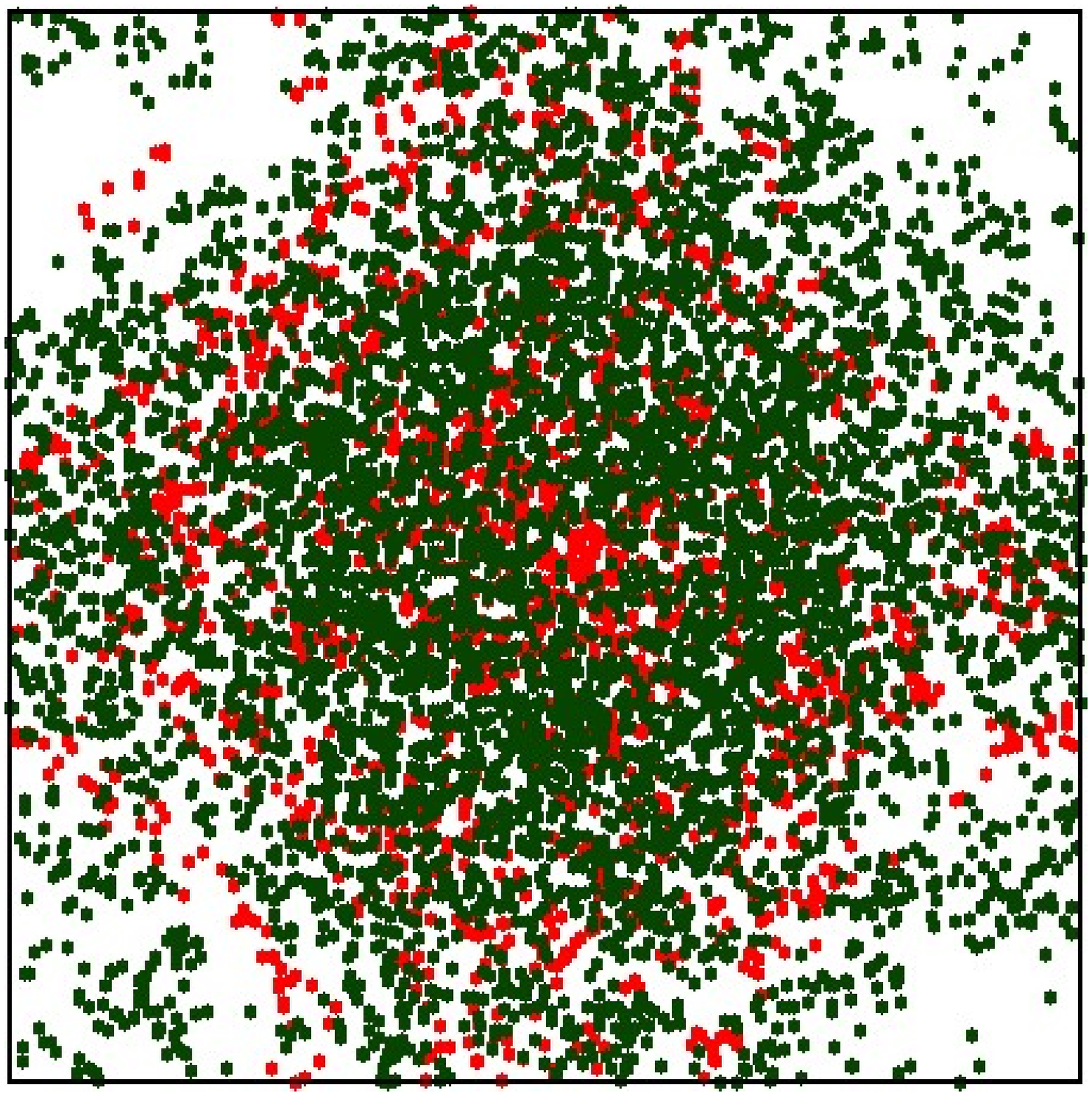}
\includegraphics[width=0.32\linewidth]{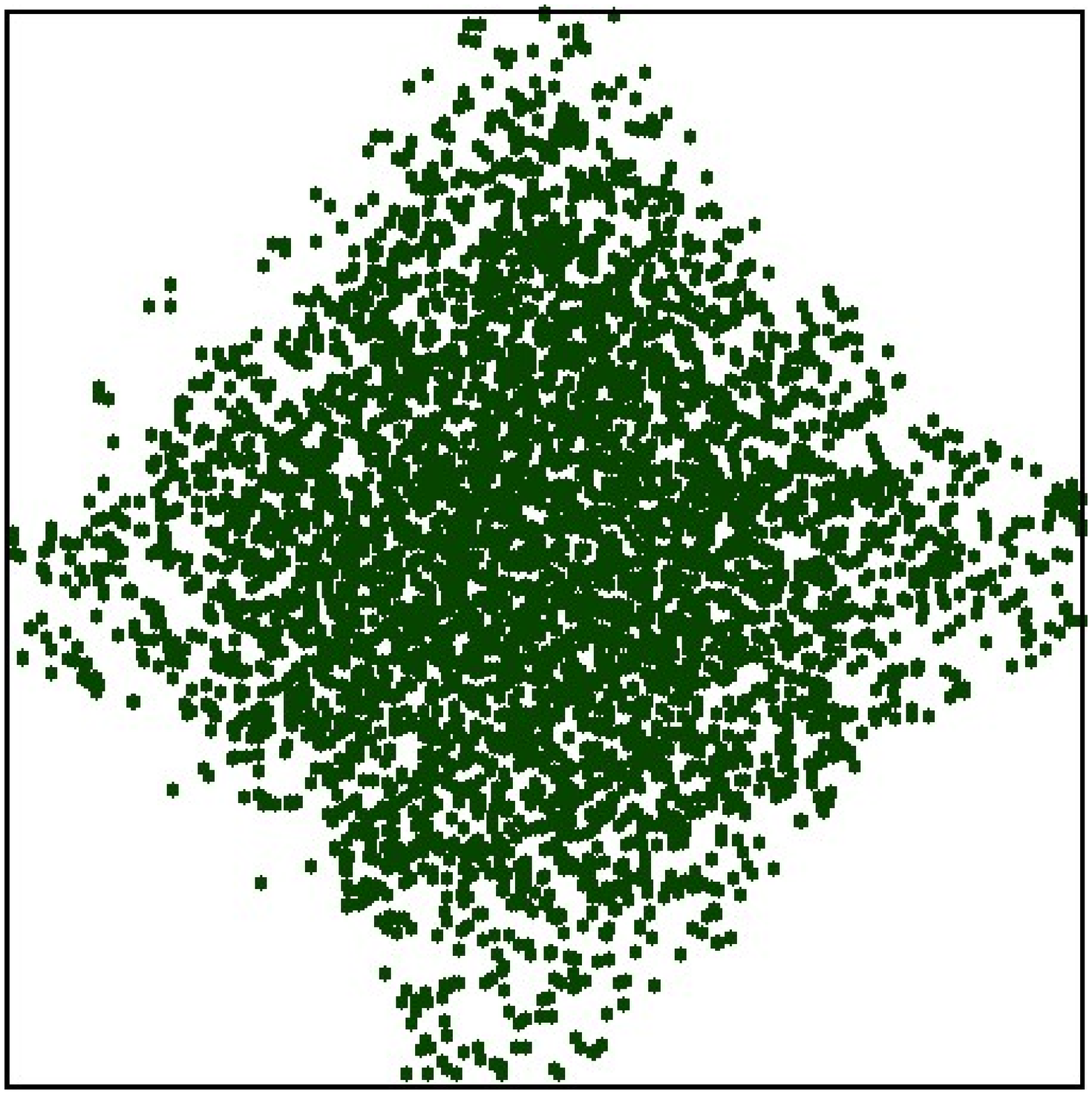}
\end{center}
\caption{\label{fig:dsnap} Competition between two neutral species
  (shown in red and green) in the steady flow with $\kappa=0.0027$. At
  time $t=0$ (left), the species are randomly distributed over the
  entire domain again at the equilibrium carrying capacity possible in
  absence of flow. Species are rotated and collapsed by the advecting
  flow towards the origin where competition takes place.  This
  progression is highlighted in the middle plot which is chosen at a
  later time $t=17$ (middle). At much later times $t=41$ (right)
  fixation occurs and only one of the species survive.}
\end{figure}

The competition between species for the two flow conditions described
above is shown in Figs.~\ref{fig:csnap} and \ref{fig:dsnap}.
Initially the populations are well-mixed at the steady state carrying
capacity as they would be with ordinary diffusion, birth, and
competitive death in absence of advection. Advection moves the
population towards the localized sinks of the flow and enhances the
competitive death embodied in the $\lambda_{ij}$ couplings. Indeed,
the middle frames of Figs.~\ref{fig:csnap} and \ref{fig:dsnap} show
explicitly the compression that leads to enhanced inter-species and
intra-species competition. Eventually at later times, only one species
survives [right hand frames of Figs.~\ref{fig:csnap} and
\ref{fig:dsnap}]. Although the extreme ($10^3$-fold!) reduction in
population size shown in Fig.~\ref{fig:dsnap} results from the use of
a maximally compressible ($\kappa=1$) turbulent flow, reductions of
$80\%$ arise for $k=0.17$ [\cite{prasad_PRL}] and even for much
smaller value of $\kappa$ [\cite{prasad_aXv}].

\begin{figure}[!h]
\includegraphics[width=0.5\linewidth]{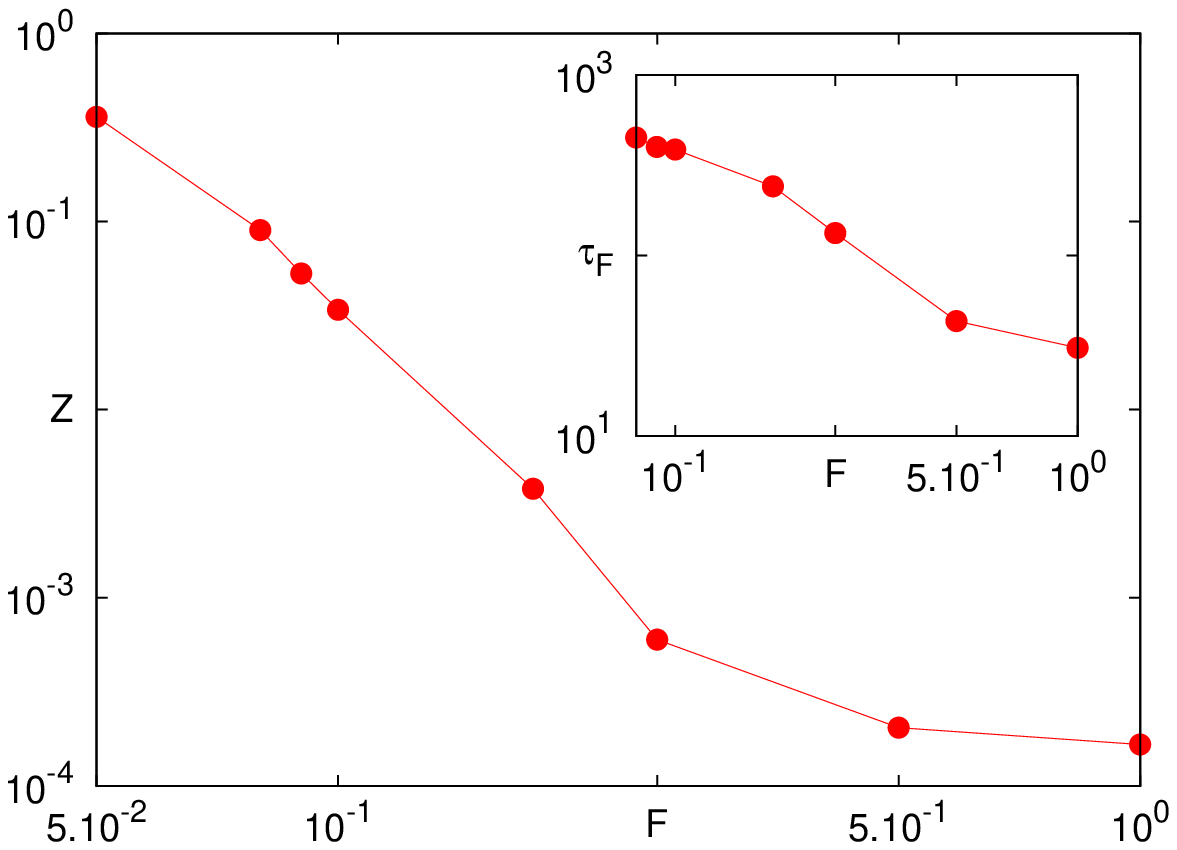}
\includegraphics[width=0.5\linewidth]{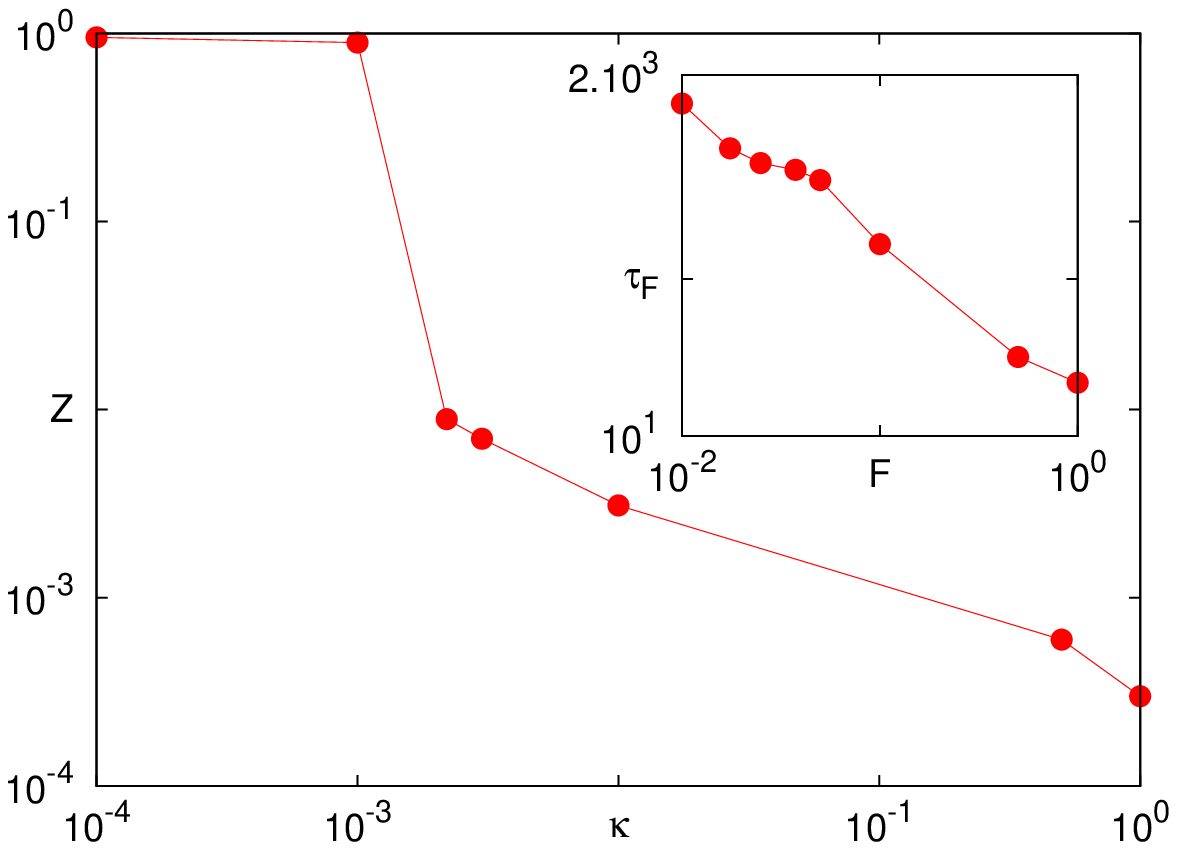} 
\caption{\label{fig:kfig1} (Left) Carrying capacity for the turbulent
  compressible flow for varying forcing strength with $\kappa=1$. $Z$
  drops with increasing forcing strength. (Right) Carrying capacity
  for the steady flow at varying compressibility levels. For very
  small compressibility, carrying capacity is close to the one in
  absence of flow and then drops. For the extreme case of $\kappa=1$,
  carrying capacity is reduced by a factor of $10^3$, similar to the
  reduction found when $\kappa=1$ for the compressible surface
  flow. In both cases, the inset reveals the drop in the fixation
  times for varying forcing at $\kappa=1$.}
\end{figure}

To quantify how advecting compressible flows affect carrying capacity
and the fixation times, we systematically vary the strength of the
flow $F$. Fig.~\ref{fig:kfig1}(left) shows that on increasing $F$, the
carrying capacity drops, due to enhanced confinement and hence
competition between the species. On the other hand, using the steady
flow we show that at fixed forcing strength, carrying capacity is also
reduced on increasing the compressibility [see
Fig.~\ref{fig:kfig1}(right)]. The insets to these figures show the
corresponding reduction in the fixation times.

\section{Conclusions}\label{conclusions}

The understanding of growth, competition, cooperation and diffusion in
space in individual based models has been subject of intense study, in
contexts as diverse as population genetics \citep{barton}, ecology
\citep{law,birch} and physics \citep{lopez,berti,korolev_RMP}. A main
focus has been to explore the regime in which discreteness effects are
such that individual based simulations differ significatively from the
behavior of macroscopic continuum equations, such as the Fisher equation or its
stochastic variant.

In this paper, we have explored competition and cooperation between
two different alleles when the total population size is not
constrained. We have deliberately focused on the weak noise limit by
choosing carrying capacities and diffusion constants such that there
is a good agreement between the outcome of the macroscopic Langevin
equations and the individual based simulations.

We have shown that, in certain limits, one can draw an explicit
correspondence with stepping stone-like models in which the total
density of individuals is kept fixed at every deme, by studying the
relative fraction of one of the two species. In the neutral case, the
fluctuating total density appear in the equation for the relative
fraction, but its fluctuations average out in the equation for mean
quantities such as the heterozygosity. The correspondence between
stepping stone models and our generalized off-lattice model with
additional fluctuations in the overall density was confirmed by
individual based simulations. In non-neutral settings, the total
density does {\em not} obey a closed equation and such exact
correspondence can not be drawn. However, we have shown how, when the
departure from neutrality is not severe (small $s$ or small
$\epsilon_A$ and $\epsilon_B$), the corrections due to density
fluctuations can be safely neglected and the predictions of
constant-density models are still reproduced with accuracy. The issue
we address here is a more subtle dynamical version of justifying the
neglect of number fluctuations in the grand canonical ensemble as
compared to the canonical ensemble in equilibrium statistical
mechanics.  We conclude that the model we present here is a natural
candidate to study situations in which the total density of
individuals can vary greatly from the background carrying capacity due
to external forces, such as turbulence or compressible fluid flows
\citep{pigo}.

{\bf Acknowledgements:} Some of the computations in this paper were
run on the Odyssey cluster supported by the FAS Science Division
Research Computing Group at Harvard University. Work by DRN was
supported through the National Science Foundation, via Grant
DMR1005289 and through the Harvard Materials Research Science and
Engineering Center, through grant DMR0820484. We thank K. Korolev for
comments on the manuscript.

\appendix
\section{Derivation of the macroscopic equations}

In this section we present an explicit derivation of the coupled
stochastic macroscopic equations for $c_A(x,t)$ and $c_B(x,t)$,
Eq.(\ref{eq_general}), from the microscopic rate reactions
(\ref{reactions}). The formalism we will follow is that of the chemical
master equation, as presented for example by \citet{gardiner}, which
in turn may be considered as a spatial generalization of the
Kramers-Moyal expansion \citep{gardiner,risken}.

As discussed in the section \ref{mod_sec}, we consider interacting
individuals in a volume equal to $L^d$ in $d$ dimensions. In
particular, competition occurs when individuals are within a small
volume $\delta$ (for details on the implementation of the
individual-based dynamics see \citet{prasad_proc}).  We can then
discretize the system in cells of size $\delta$ and start the
derivation from the master equation governing the time evolution of
the probability of the numbers of particles $\{n_j^A,n_j^B\}$ of type $A$
and $B$ in each cell, labeled by the index $j$. We first define as
$W_A(\pm 1,n_j^A,n_j^B)$ and $W_B(\pm 1,n_j^A,n_j^B)$ the rates at
which the populations of type $A$ (or $B$) increase/decrease by one
individual in a specific box, given that the numbers are currently
$n^A_j$ and $n^B_j$. The expression for these rates are then:

\begin{eqnarray}\label{mrates}
  W_A(+1,n_j^A,n_j^B)&=&\mu_A n_j^A\nonumber\\
  W_A(-1,n_j^A,n_j^B)&=&\tilde{\lambda}_{AA}n_j^A(n_j^A-1)+\tilde{\lambda}_{AB}n_j^An_j^B\nonumber\\
  W_B(+1,n_j^A,n_j^B)&=&\mu_A n_j^B\nonumber\\
  W_B(-1,n_j^A,n_j^B)&=&\tilde{\lambda}_{BA}n_j^An_j^B+\tilde{\lambda}_{BB}n_j^B(n_j^B-1).
\end{eqnarray}

The master equation governing the evolution of the full probability
distribution $P(\{n_j^A,n_j^B\},t)$ for all possible box occupation
numbers $\{n^A_j,n^B_j\}$ then reads:
\begin{eqnarray}\label{generalmastereq}
 \frac{d}{dt}P(\{n_j^A,n_j^B\},t)&=&\sum_j\left[W_A(+1,n_j^A-1,n_j^B)P(n_1^A,\dots,n_j^A-1,\dots,n_1^B,\dots)
-W_A(+1,n_j^A,n_j^B)P(\{n_j^A,n_j^B\})\right]\nonumber\\
&+&  \sum_j\left[W_A(-1,n_j^A+1,n_j^B)P(n_1^A,\dots,n_j^A+1,\dots,n_1^B,\dots)
-W_A(-1,n_j^A,n_j^B)P(\{n_j^A,n_j^B\})\right]\nonumber\\
&+& \sum_j\left[W_B(+1,n_j^A,n_j^B-1)P(n_1^A,\dots,n_1^B,\dots,n_j^B-1,\dots)
-W_B(+1,n_j^A,n_j^B)P(\{n_j^A,n_j^B\})\right]\nonumber\\
&+& \sum_j\left[W_B(-1,n_j^A,n_j^B+1)P(n_1^A,\dots,n_1^B,\dots,n_j^B+1,\dots)
-W_B(-1,n_j^A,n_j^B)P(\{n_j^A,n_j^B\})\right]\nonumber\\
&+& \mathrm{diffusion}\ \mathrm{terms},
\end{eqnarray}

\noindent where the diffusion terms allow for the stochastic exchange
of particles between neighboring boxes.  Although we did not write
them explicitly, they reduce to discrete approximations to Laplace
operator. Indeed, we replace them with Laplacians in the continuous
space limit at the end of the calculation.

The next step in the derivation is to perform a Kramers-Moyal
expansion \citep{risken} in each of the boxes, which leads to

\begin{equation}\label{KM1}
  \partial_t P(\{n_j^A,n_j^B\})=\sum\limits_j\sum\limits_{k=1}^{\infty} 
\frac{(-1)^k}{k!}\{\partial^k_{n_j^A}[\alpha^A_k(n_j^A,n_j^B)P(\{n_j^A,n_j^B\})]+\partial^k_{n_j^B}
[\alpha^B_k(n_j^A,n_j^B)P(\{n_j^A,n_j^B\})]\},
\end{equation}
with
\begin{equation}\label{KM2}
\alpha_k^{A,B}(n_j^A,n_j^B)=\int d\Delta n_j^{A,B} \ (\Delta n_j^{A,B})^k W_{A,B}(\Delta n_j^{A,B},n_j^A,n_j^B),
\end{equation}
and where the integral over $\Delta n$ accounts for the possible jump
processes ($+1$ and $-1$ in our case).  Finally, truncating the
Kramers-Moyal expansion up to second order in the derivatives leads to
a Fokker-Planck equation for $P\{n_j^A,n_j^B\}$. It is convenient to
write directly the equivalent but somewhat simpler system of Langevin
equations corresponding to this Fokker-Planck description, namely:
\begin{eqnarray}\label{eqnumbdens}
 \frac{dn_j^A}{dt}=n_j^A(\mu_A-\tilde{\lambda}_{AA}n_j^A-\tilde{\lambda}_{AB}n_j^B)
+\mathrm{diffusion}
  +\sqrt{n_j^A(\mu_A+\tilde{\lambda}_{AA}n_j^A+\tilde{\lambda}_{AB}n_j^B)}\xi_j^A
  \nonumber\\
  \frac{dn_j^B}{dt}=n_j^B(\mu_B-\tilde{\lambda}_{BA}n_j^A-\tilde{\lambda}_{BB}n_j^B)
+\mathrm{ diffusion}
  +\sqrt{n_j^B(\mu_B+\tilde{\lambda}_{BA}n_j^A+\tilde{\lambda}_{BB}n_j^B)}\xi_j^B.
\end{eqnarray}

In the above system of equations, the $\xi$'s are delta-correlated
unit variance Gaussian processes, $<\xi_j^k(t)
\xi_l^m(t')>=\delta_{jl}\delta_{km}\delta(t-t')$. The multiplicative
noise in the equation must be interpreted according to the Ito
prescription \citep{gardiner,korolev_RMP}. In principle,
the diffusion terms in (\ref{generalmastereq}) would contribute to the
noise term. However, one can show that this contribution can be
neglected if the size of the cells is sufficiently large (see
\citet{gardiner}).

From Eqs.(\ref{eqnumbdens}) one can finally derive
Eqs. (\ref{eq_general}) by:
\begin{enumerate}
\item Taking (formally) the limit $\delta\rightarrow 0$. In such a way
  the number densities of individuals are continuous functions of
  the coordinate $\mathbf{x}$, $n_{A}(\mathbf{x},t)$ and $n_{B}(\mathbf{x},t)$.
\item Defining rescaled, macroscopic rates of binary reactions,
  \begin{equation}\label{map_param}\lambda_{ij}=N\delta\tilde{\lambda}_{ij}\end{equation}
\item Defining the macroscopic concentrations of individuals  $c_{A,B}(\mathbf{x},t)=n_{A,B}(\mathbf{x},t)/N$.
\end{enumerate}

The convenience of introducing the macroscopic binary reaction rates
$\lambda_{ij}$ in step (2) is that the microscopic interaction radius
$\delta$ does not appear in the macroscopic system of equations
(\ref{eq_general}). At the same time, we introduced a parameter
$N=\lambda_{ij}/(\delta\tilde{\lambda}_{ij})$ that, as clear from step
(3) in the above procedure, sets the typical number density of
particles corresponding to a macroscopic concentration
$c(\mathbf{x},t)=1$. Such parameter does not appear in the
deterministic drift terms of the equation but only in the noise terms,
whose amplitude vanishes for $N\rightarrow\infty$.  It is worthwhile
remarking that, while we followed here the Kramers-Moyal expansion
procedure, in the Van Kampen formalism the parameter $N^{-1}$ is the
relevant expansion parameter which is assumed to be small
\citep{risken,gardiner}.

We remark that through the paper we presented only results of the
particle models, corresponding to given parameter choices in the
macroscopic equations (\ref{eq_general}). Equation (\ref{map_param})
can be seen as defining the mapping between the parameters used in the
particle simulations (the interaction domain $\delta$ and the
microscopic binary rates $\tilde{\lambda}_{ij}$'s) and those appearing
in the macroscopic description ($N$ and the set of $\lambda_{ij}$'s)
The same relation can be used for the reverse task, i.e. finding
microscopic parameters $\delta$ and $\tilde{\lambda}_{ij}$'s
corresponding to given $N$ and $\lambda_{ij}$'s. Clearly this mapping
is not univoquely determined, but has one degree of freedom. As
sketched in Sec. (\ref{mod_sec}), we fixed this degree of freedom in
two different ways in the well-mixed version of the model and in the
$d>0$, spatially explicit simulations. In particular, in $d=0$ we
chose $\delta=1$, so that the microscopic binary reaction rates are
$N$ times smaller than the macroscopic ones,
$\tilde{\lambda}_{ij}=\lambda_{ij}/N$. In this case, it is crucial to
set the system size $L=1$ so that all particles interact with all
other particles. Instead, in $d>0$ we chose the interaction domain
$\delta=1/N$, so that the microscopic and macroscopic reaction rates
are identical, $\tilde{\lambda}_{ij}=\lambda_{ij}$. Further details on
the simulation schemes can be found in \citet{prasad_proc}.

We conclude this Appendix by noting that the continuous space limit
is a formal one, and cannot be performed in a rigorous way.  One of
several subtleties is that neglect of the diffusive contribution to
the noise variance requires a finite value of $\delta$, so that the
limit of vanishingly small interaction range cannot be taken in a
strict sense.  Thus, Eq. (\ref{generalmastereq}) should be regarded as
a continuum shorthand notation: In practice, we always simulate
equations such as Eq. (\ref{eqnumbdens}) on a lattice of finite size,
and require a smoothly varying total density of particles. When this
assumption is invalid, the macroscopic description can break down, as
briefly discussed in the beginning of section (\ref{section_space})
for the problem of the reduction in the total number of particles for
$d=1$ and $d=2$.

\section{Appendix: equations for the relative fraction of one species}
\label{app_f}


The correspondence between the growth model presented here and the
stepping stone model with Fisher-Wright or Moran dynamics, or the
equivalent stochastic Fisher-Kolmogorov-Petrovsky-Piscounov equation
\citep{fisher,kolmogorov} can be illuminated by constructing the dynamical
equation for the relative fraction of species $A$, $f=c_A/c_T$ with
$c_T=c_A+c_B$. A dynamical equation for $f$ can be derived with help
of the Ito calculus: upon writing the system of equation
(\ref{eq_general}) as:
\begin{eqnarray}
\frac{d}{dt} c_A (\mathbf{x},t)& = &\alpha_A(c_A,c_B) + 
\sigma_A(c_A,c_B)\xi (\mathbf{x},t) \nonumber\\
\frac{d}{dt} c_B(\mathbf{x},t) & = &\alpha_B(c_A,c_B) + 
\sigma_B(c_A,c_B)\xi'(\mathbf{x},t)
\end{eqnarray}
where the diffusive Laplacian terms are included into  $\alpha_A$, $\alpha_B$.
The equation for the $A$-fraction $f$ then reads
\begin{eqnarray}\label{itoform}
\frac{d}{dt}f &=& \alpha_A\partial_Af+\alpha_B\partial_Bf
+\sqrt{\sigma_A^2(\partial_Af)^2+\sigma_B^2(\partial_Bf)^2}\xi + 
\nonumber\\
&+&\frac{\sigma^2_A}{2}\partial_{AA}f+\frac{\sigma^2_B}{2}\partial_{BB}f,
\end{eqnarray}
where we used the abbreviated notation
$\partial_A\equiv\partial_{c_A}$,
$\partial_{AA}\equiv \partial^2_{c_A}$ and so on. Inserting the
complete set of equations \ref{eq_general} into (\ref{itoform}) leads
to a lengthy expression for the dynamics of$f$. However, with the
choice of parameters we made to discuss a reproductive advantage (this
reduces to the neutral case for $s=0$), Eq.(\ref{itoform}) simplifies
to
\begin{eqnarray}
\frac{\partial}{\partial t}f &=&D\nabla^2f +2D\nabla(\log c_T)\cdot \nabla f + \nonumber\\
&+&\mu s f(1-f) + \frac{\mu s f}{c_TN}(f-1)
+\sqrt{\mu f(1-f)\frac{1+c_T}{Nc_T}+\frac{\mu s f}{N c_T}(1-f)^2}\xi. 
\end{eqnarray}
Upon neglecting small contributions of order $s/N\ll 1$ in the last
two terms and neglecting fluctuations in the total density
(i.e. imposing $c_T=1$), we recover exactly the equation
(\ref{stochasticfisher}) governing the macroscopic dynamics of the
stepping stone model.

Repeating the calculation in the case of mutualism yields:

\begin{eqnarray}\label{mut_eq_mf}
  \frac{\partial}{\partial t}f &=&D\nabla^2f +2D\nabla(\log c_T)\cdot \nabla f + 
\mu f(1-f)[\epsilon_A-(\epsilon_A+\epsilon_B )f]+\nonumber\\
&+& 
  \frac{\mu f(f-1)}{N}[\epsilon_A(f-1)+\epsilon_Bf]
  +\sqrt{\frac{\mu f(1-f)\left[\left(\frac{1+c_T}{c_T}\right)
        -\epsilon_A(1-f)^2-\epsilon_Bf^2 \right]}{N}}\xi
\end{eqnarray}
Upon neglecting, similar to the case of reproductive advantage, terms
order $\epsilon_{A,B}/N$, and again neglecting fluctuations away from
the line
$c_T(\mathbf{x},t)=c_A(\mathbf{x},t)+c_B(\mathbf{x},t)=1$, we recover
the continuum limit of the mutualistic stepping stone model treated by
\citet{korolev_mut}, namely
\begin{equation}
  \frac{\partial}{\partial t}f = D\nabla^2f + s_0f(1-f)(f^*-f)+\sqrt{\frac{2\mu f (1-f)}{N}}\xi(\mathbf{x},t),
\end{equation}
where $s_0=\mu(\epsilon_A+\epsilon_B)$ and $f^*=\epsilon_A/(\epsilon_A+\epsilon_B)$.

\section{Appendix: Fixation times for the mutualistic model in the well-mixed case}
\label{app_mut_mf}

To estimate the average fixation time for the mutualistic model in the
well-mixed limit, we start from Eq. \ref{mut_eq_mf}. Upon neglecting
terms order $\epsilon_A/N$, $\epsilon_B/N$ and also neglecting density
fluctuations by imposing $c_T=c_A+c_B=1$, we obtain:

\begin{equation}\label{mut_kir}
\frac{d}{dt}f \approx \mu f(1-f)[\epsilon_A-(\epsilon_A+\epsilon_B )f]
+\sqrt{\frac{2\mu f(1-f)}{N}}\xi.
\end{equation}

The dynamics of such equation will reach one of the two absorbing
states at $f=0$ or $f=1$ for long enough times. However, these times
can be very long when $N$ is large: a time-independent metastable
probability distribution exists before the absorbing states are
reached, which can be written using potential methods \citep{gardiner} as
\begin{equation}
\mathcal{P}(f)\propto e^{-V(f)}
\end{equation}
where the potential $V$ is given by
\begin{equation}
V(f)=-Nf\left[\epsilon_A-\frac{(\epsilon_A+\epsilon_B)}{2}f\right]
+\ln[f(1-f)]
\end{equation}
where the first term is analogous to a potential energy and the second
resembles an entropy.  In the large $N$ limit, the potential has a
minimum at $f^c\approx \epsilon_A/(\epsilon_A+\epsilon_B)$ and two
maxima, one at $f^-\approx 1/(N\epsilon_A)$ and one at $f^+\approx
1-1/(N\epsilon_B)$. By evaluating the potential at these points one
can estimate the lifetime of the metastable state from the height of
the two potential barriers.  To the leading order in $N$, the smallest
barrier is given by:
\begin{equation}
\Delta V= \frac{N}{2}\frac{\min(\epsilon_A^2,\epsilon_B^2)}
{\epsilon_A+\epsilon_B}.
\end{equation}
Finally, we assume that fixation always occurs via the smallest
barrier. With this assumption, the time needed to escape the potential
minimum to one of the absorbing state can be simply estimated from
Kramer's escape rate theory as $t^*\sim \exp(\Delta V)$, which leads to Eq.(\ref{kramer}).

We now discuss the fixation probability in zero dimensions. The Kolmogorov
backward equation corresponding to the stochastic differential equation
(\ref{mut_kir}), when interpreted using the Ito calculus, reads:

\begin{equation}
\frac{\partial u(p,t)}{\partial t} = \frac{1}{N}p(1-p)\frac{\partial^2}{\partial p^2}u(p,t)
~+\tilde{s}p(1-p)(f^* -p) \frac{\partial}{\partial p}u(p,t),
\label{backwards}
\end{equation}
where $u(p,t)$ is the probability that species A has fixed at time $t
> 0$ given that it was present with frequency $p$ at time $t=0$.  We
have set $f^*=\epsilon_A/(\epsilon_A+\epsilon_B)$, and defined the
{\em mutualistic} advantage $\tilde{s}=\mu(\epsilon_A+\epsilon_B)$.

Note that Eq. \ref{backwards} includes the original Kimura problem of
two non-interacting species as a special case, in the limit $f^* \to
\infty, \tilde{s} \to 0$ with the selective advantage given by the fixed
product, $ s \equiv f^* \tilde{s}\equiv \mu \epsilon_A$.  
We now define the long time
fixation probability for the initial condition $p=f_0$ as
\begin{equation}
\lim_{t \to \infty} u(f_0,t) \equiv u(f_0)
\end{equation}
Upon assuming a steady state arises at long times, we have from Eq. (\ref{backwards})
\begin{equation}
\frac{d}{dp} u'(p) ~=~ Ns(f^* -p) u'(p)
\end{equation}
which leads to
\begin{equation}
u'(p) ~=~ C~e^{\frac{1}{2}Ns(f^* -p)^2}
\end{equation}
With boundary conditions $u(0)=0,~u(1)=1$, we integrate once more to obtain
the fixation probability \citep{korolev_mut}
\begin{equation}
u(f_0) ~=~ \frac{\int_0^{f_0}e^{\frac{1}{2}Ns(f^* -p)^2}dp}{\int_0^{1}e^{\frac{1}{2}Ns(f^* -p)^2}dp}~~,
\label{fixation}
\end{equation}
a closed form expression in terms of the parameters $f_0,f^*,N ~{\rm
  and}~ s$. It is straightforward to show that in the limit $f^* \to
0, s \to 0$ with $\tilde s \equiv f^* s$ fixed (two noninteracting
species with a selective advantage $\tilde s$) we recover Kimura's
famous formula for the fixation probability, Eq. (\ref{Kimur}).

\bibliographystyle{elsarticle-harv}
\bibliography{noflowpaper}







\end{document}